\documentclass[%
superscriptaddress,
 amsmath,amssymb,
 aps,
twocolumn,
english
]{revtex4-2}

\usepackage{graphicx}
\usepackage{dcolumn}
\usepackage{bm}
\usepackage[dvipsnames]{xcolor}

\begin{document}

\title{Photon control and coherent interactions via lattice dark states in atomic arrays}

\author{Oriol Rubies-Bigorda}
\email{orubies@mit.edu}
\affiliation{Physics Department, Massachusetts Institute of Technology, Cambridge, Massachusetts 02139, USA}
\affiliation{Department of Physics, Harvard University, Cambridge, Massachusetts 02138, USA}
\author{Valentin Walther}
\affiliation{ITAMP, Harvard-Smithsonian Center for Astrophysics, Cambridge, Massachusetts 02138, USA}
\affiliation{Department of Physics, Harvard University, Cambridge, Massachusetts 02138, USA}
\author{Taylor L. Patti}
\affiliation{Department of Physics, Harvard University, Cambridge, Massachusetts 02138, USA}
\author{Susanne F. Yelin}
\affiliation{Department of Physics, Harvard University, Cambridge, Massachusetts 02138, USA}


\begin{abstract}
Ordered atomic arrays with subwavelength spacing have emerged as an efficient and versatile light-matter interface, where collective interactions give rise to sets of super- and subradiant lattice states. Here, we demonstrate that highly subradiant states, so-called lattice dark states, can be individually addressed and manipulated by applying a spatial modulation of the atomic detuning. More specifically, we show that lattice dark states can be used to store and retrieve single photons with near-unit efficiency, as well as to control the temporal, frequency and spatial degrees of freedom of the emitted electromagnetic field. Furthermore, we demonstrate how to engineer arbitrary coherent interactions between multiple dark states and thereby manipulate information stored in the lattice. These results pave the way towards quantum optics and information processing using atomic arrays.
\end{abstract}

\maketitle


\section{\label{sec:introduction}Introduction}

Ordered atomic arrays with subwavelength spacing have emerged as 
versatile quantum many-body systems, where coherent excitation exchange between the dipoles leads to a collective response of the atomic ensemble \cite{Asenjo_2017_general,Ephi_2017,PRL_CAdams}. The richness of the underlying interactions can be used for a wide variety of quantum applications that range from topological phases of matter \cite{Janos_2017_PRL,Janos_2017,Topological_Adams}, atomic clocks \cite{clock_1,clock_2} and optical quantum memories \cite{Manzoni_2018} to the ability to modify the radiative environment of single impurities \cite{Taylor_controllinginteractions,Ana_controllinginteractions}. Atomic arrays also offer remarkable optical properties and 
can act as an optical mirror for incident beams of low intensities \cite{Ephi_2017,Bloch_mirror,Ruostekoski_steering}. Additionally, this platform could eventually be used to manipulate quantum light, create cat and photonic Greenberger-Horne-Zeilinger states suitable for quantum information processing \cite{Bekenstein2020} and build photonic quantum gates by exploiting Rydberg interactions in the few-photon limit \cite{morenocardoner_twophotons,valentin_rydberg}.

Central to understanding the physics of atomic arrays is the notion that excitations in the lattice are characterized by their momentum \cite{Asenjo_2017_general}. 
For lattice spacings smaller than the atomic transition wavelength, a set of lattice excitations emerges whose momenta are larger than that of any resonant electromagnetic field mode.
Such excitations do not radiate and are therefore called subradiant or dark states. First proposed by Dicke \cite{Dicke,GrossHaroche}, subradiant states have been thoroughly studied in various systems \cite{subradiance_measurement_1,subradiance_measurement_2,subradiance_measurement_3,Browaeys_subradiance,storage_2,storage_3} due to their long life-times and their potential applications in sensing \cite{sensing_1,sensing_2}. While subradiant states have recently been observed in two-dimensional atomic arrays \cite{Bloch_mirror}, they can only be accessed with great difficulty due to the fact that they do not couple to incoming light fields. One can circumvent this problem by placing an impurity close to the array \cite{Ritsch_impurity,Ana_controllinginteractions,Taylor_controllinginteractions}, by exploiting the Zeeman splitting between $J=1$ atomic levels to access a small subset of the subradiant states that emerge in two-dimensional arrays \cite{storage_1} or by applying magnetic field gradients to imprint different phases on each atom \cite{Jen_single}.

\begin{figure}[b!]
\includegraphics[width=\linewidth]{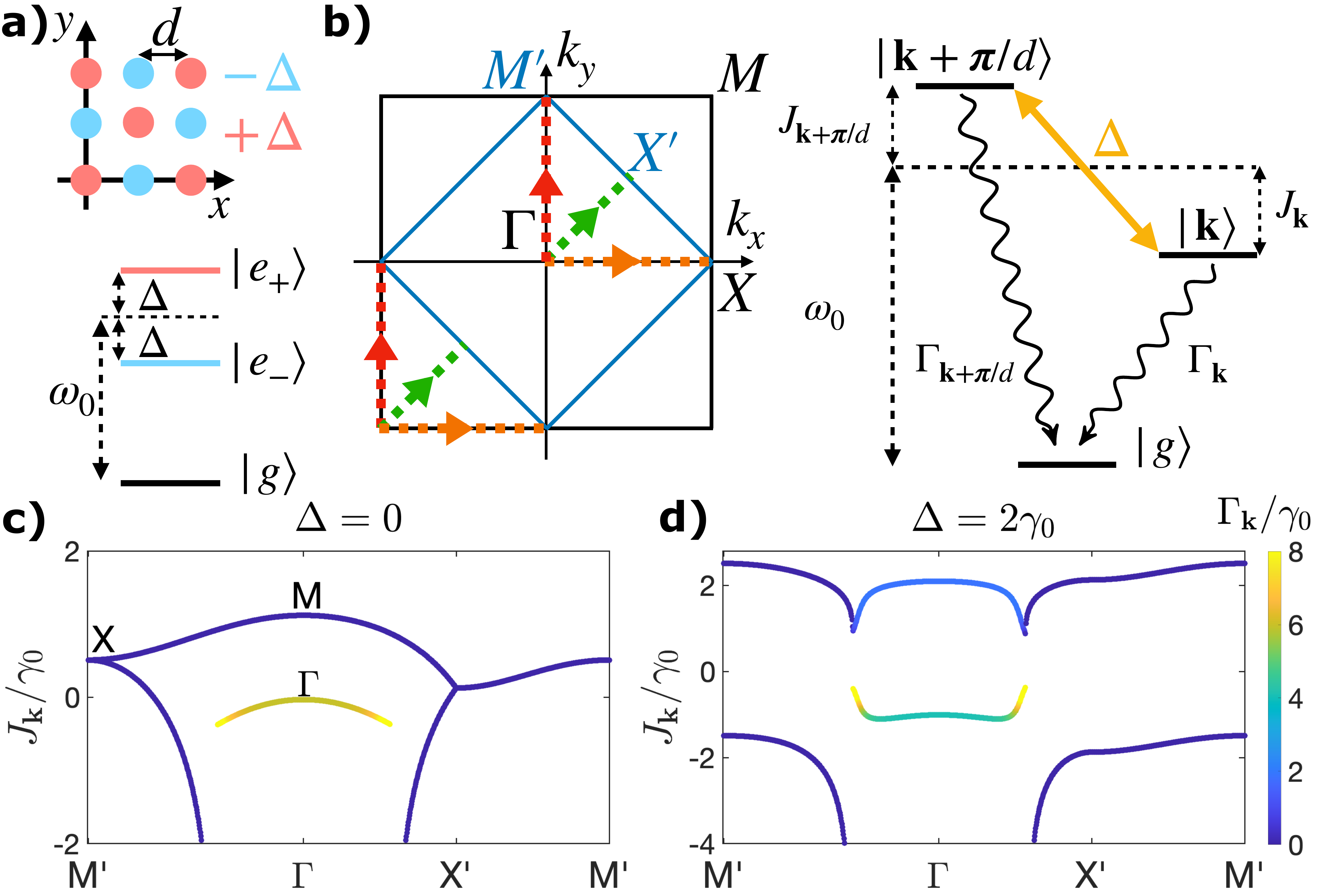}
\caption{\label{fig: schematic} \textbf{Checkerboard detuning pattern.} (a) The checkerboard lattice is a non-Bravais lattice with two atoms per unit cell, such that neighboring atoms have detunings of opposite sign relative to the natural transition frequency $\omega_0$. (b) Momentum space representation of a square lattice (black square) and the checkerboard lattice that emerges at finite detuning (blue square). The detuning couples the momentum states $\mathbf{k}$ and $\mathbf{k}+\boldsymbol{\pi}/d$ (dashed lines) with strength $\Delta$ and creates a three level system such that both momentum states decay to the ground state with their respective collective decay rates. (c) Band structure 
for the nondetuned lattice with spacing $d=0.2\lambda_0$ and circular in-plane polarization. 
The radiating states are located around the origin of the Brillouin zone $\Gamma$ and zero decay is observed outside the lightcone. (d) Upon introducing a finite detuning, a bandgap opens. The momentum states around $\Gamma$ mix with those close to the corner $M$ and the original dark band acquires a nonzero decay rate. }
\end{figure}

Interestingly, coupling of subradiant modes has long been achieved in other physical systems by spatially modulating one of its elements or parameters. For example, subwavelength gratings have allowed imaging beyond the diffraction limit by converting evanescent waves near a surface into propagating waves \cite{Liu_superlens}. A similar technique was used to couple propagating light to subradiant surface plasmonic modes by means of a grating on a metallic tip \cite{Ropers_Plasmon_grating}. Additionally, spatial modulations of the refractive index have been leveraged to access evanescent modes of subwavelength photonic lattices \cite{Alfassi_modulation_index_refraction}. Recent theoretical studies have applied these ideas to atomic arrays in order to access topological boundary states that exhibit subradiance \cite{Zhang_1D_detuningpattern} and to study quantum phase transitions \cite{Ruostekoski_nature}.
In this case, the symmetry of the atomic lattice is broken by applying a spatial modulation of the atomic detuning, which  couples radiating states to dark states and therefore provides a handle to redistribute excitations between the two. Intuitively, this detuning landscape introduces a new lengthscale larger than the lattice spacing. Given that subradiant modes can only be sustained if the new lengthscale is smaller than a wavelength, lattice excitations that were originally dark are now forced to radiate. Conversely, these states can be populated by an incoming light field while the detuning pattern is present. Upon turning the detuning off, they recover their long-lived nature and remain in the array.

Here, we prove that spatial detuning patterns allow to selectively address individual lattice dark states. Also, we demonstrate that this technique can be used to store single photons in the collective dark states of two-dimensional atomic arrays and that the excitation can be subsequently retrieved with high fidelity, even for systems with only a few hundred atoms \footnote{During the final stages of the preparation of this manuscript, we became aware of a recent arxiv submission by Ballantine and Ruostekoski \cite{roustekoski_2021} with a similar proposal, where ac Stark shifts of the atomic levels are used to access subradiant modes of atomic arrays. The work discusses how to engineer a Huygens’ surface with a rectangular bilayer lattice and how to generate entanglement between the lattice and a cavity.} \cite{roustekoski_2021}. By properly choosing the magnitude of the detuning during the retrieval process, one can obtain full control over the time-frequency degrees of freedom of the emitted electric field. In particular, we show how to emit photons with an arbitrary temporal shape and how to produce single photons that are in a coherent superposition of two frequencies and can therefore be used as quantum bits \cite{controlsinglephoton_2,multidimensionalphotoncontrol_3}. Additionally, we explore different schemes to modify the spatial properties of the emitted electric field and demonstrate that beam-steering \cite{beamsteering,Ruostekoski_steering} at the single-photon level can be achieved. Interestingly, spatial modulations of the detuning do not only allow to create radiative paths from dark momentum states to the ground state, but also enable to couple multiple dark states with one another. As a result, one can engineer a wide variety of long-lived, coherent interactions between them that can potentially be used to experimentally measure the bandstructure of the system or to devise quantum gates \cite{Taylor_controllinginteractions,all_optical_1,all_optical_2}. 

\section{\label{sec:model}  Model}

Let us consider a two-dimensional square array of two-level atoms interacting with the vacuum electromagnetic field. The Hamiltonian governing the system is \cite{Lehmberg_1970_1,Lehmberg_1970_2,CohenTanoudgi_book,Meystre_book}

\begin{align}
\label{eq: FullHamiltonian}
\hat{H} & = \sum_j \hbar \left( \omega_0 -\Delta_j \right) \hat{\sigma}_j^\dagger \hat{\sigma}_j + \sum_{\boldsymbol{\kappa},\boldsymbol{\epsilon} \perp \boldsymbol{\kappa}} \hbar \omega_{\boldsymbol{\kappa}} \hat{a}_{\boldsymbol{\kappa},\boldsymbol{\epsilon}}^\dagger \hat{a}_{\boldsymbol{\kappa},\boldsymbol{\epsilon}} +\hat{V}.
\end{align} 

\noindent Here, the index $j$ labels the atom at position $\mathbf{r}_j$ and $\Delta_j$ represents the detuning of that atom with respect to its bare frequency $\omega_0$. $\boldsymbol{\kappa}$ is the three-dimensional wavevector of a given electromagnetic mode and $\boldsymbol{\epsilon}$ its polarization. The last term, $\hat{V}$, describes the interaction between the radiation field and the lattice atoms and depends on the atomic dipole moment and the vacuum electric field at the atomic positions

\begin{equation}
    \hat{V}=\sum_j \hat{\mathbf{P}}_j \hat{\mathbf{E}}(\mathbf{r}_j) = -\hbar \sum_{j} \sum_{\boldsymbol{\kappa},\boldsymbol{\epsilon} \perp \boldsymbol{\kappa}}  g_{\boldsymbol{\kappa}} \left( \mathbf{d} \boldsymbol{\epsilon} \hat{\sigma}^\dagger_j \hat{a}_{\boldsymbol{\kappa},\boldsymbol{\epsilon}} e^{i \boldsymbol{\kappa} \mathbf{r}_j} +H.c. \right), 
\end{equation}

\noindent where $g_{\boldsymbol{\kappa}}=d_0 \sqrt{\omega_{\boldsymbol{\kappa}}/2\hbar\epsilon_0V}$ and the direction and magnitude of the dipole moment $\mathbf{d}$ are assumed to be the same for all atoms.

\subsection{Lattice dynamics}

Applying the Born-Markov approximation, one can trace out the electromagnetic degrees of freedom and obtain the master equation describing the density operator of the atoms \cite{Lehmberg_1970_1,Lehmberg_1970_2,CohenTanoudgi_book}. Equivalently, the dynamics of the system are described by an effective non-Hermitian Hamiltonian $\hat{H}_{\text{eff}}$ in the quantum jump formalism \cite{Dalibard_jumps_1992,Carmichael_book,Dalibard_jump_1993,Carmichael_2000}. If the lattice contains one or zero excitations (one-excitation manifold) and no external drive is applied, the quantum jumps have no effect and the lattice dynamics are fully determined by $\hat{H}_{\text{eff}}$ \cite{Asenjo_2017_general,Janos_2017}

\begin{align}
\label{eq: efectiveHamiltonian}
\hat{H}_{\text{eff}} = \hbar \sum_j \left( \omega_0 -\Delta_j(t) \right) \hat{\sigma}^\dagger_j \hat{\sigma}_j +\hbar \sum_{j,i}  \left(J_{ji}-i\frac{\Gamma_{ji}}{2} \right) \sigma^\dagger_j \sigma_i,
\end{align} 

\noindent where $J_{ji}$ and $\Gamma_{ji}$ are the cooperative energy shifts and decay rates arising from dipole-dipole interactions between atoms $j$ and $i$, and are given by the dyadic Green's function in free space \cite{Chew_dyadicGreens,dyadic_novotny_hecht_2006,Asenjo_2017_general}

\begin{equation}
\label{eq: shifts_greensfunction}
    J_{ji}-i\frac{\Gamma_{ji}}{2}=- \frac{3\pi \gamma_0}{\omega_0} \mathbf{d}^*  \cdot \mathbf{G}\left( \mathbf{r}_j -\mathbf{r}_i,\omega_0 \right) \cdot \mathbf{d}.
\end{equation}

\noindent Note that we define $\Gamma_{jj}=\gamma_0$ to be the spontaneous emission rate and include the Lamb shift $J_{jj}$ in $\omega_0$.
In the last equation, we neglected the dispersion of the Green's function consistent with the Markovian approximation.
Then, the field scattered by the atoms is \cite{Welsch_2002_field,Welsch_2007_field,Fan_2015_field,Caneva_2015_field}

\begin{equation}
\label{eq: electricfield_operator}
    \hat{\mathbf{E}}(\mathbf{r})= \mu_0 k_0^2 \sum_j \mathbf{G}\left( \mathbf{r}_j -\mathbf{r},\omega_0 \right) \mathbf{d} \hat{\sigma}_j.
\end{equation}

The state of the lattice in the one-excitation manifold can be expressed as $|\Psi (t) \rangle = \sum_j e_j(t) e^{-i \omega_0 t} |e_j\rangle + g(t) |g \rangle$, where $|g \rangle$ is the state where all atoms are in the ground state and $|e_j\rangle$ represents the state in which only atom $j$ is excited. Applying Schr\"{o}dinger's equation, we obtain the following equations of motion

\begin{equation}
        \frac{d e_j}{dt}= i \Delta_j(t) e_j - i \sum_{i} \left( J_{ji} -i \frac{\Gamma_{ji}}{2} \right) e_i.
\end{equation}

For an infinite lattice, one can simplify the problem by applying Bloch's theorem. Then, the momentum states of the atomic lattice can be expressed as $v_{\mathbf{k}}=\sum_j  e^{-i \mathbf{k} \mathbf{r}_j} e_j$, where $\mathbf{k}$ is a two-dimensional momentum contained within the first Brillouin zone of the reciprocal lattice. For a square lattice with interparticle spacing $d$, each component is contained within $\{-\pi/d,\pi/d\}$. The equations of motion in momentum space take the form

\begin{equation}
\label{eq: EquationMotionMomentumPrior}
        \frac{d v_{\mathbf{k}}}{dt}=  - i \left( J_{\mathbf{k}} - i \frac{\Gamma_\mathbf{k}}{2} \right) v_{\mathbf{k}} + i \sum_j e^{-i \mathbf{k} \mathbf{r}_j} \Delta_j(t) e_j,
\end{equation}

\noindent where the cooperative shifts $J_\mathbf{k}$ and decays $\Gamma_\mathbf{k}$ are given by the Fourier transform of the Green's function $\tilde{\mathbf{G}}(\mathbf{k})=\sum_j e^{-i \mathbf{k} \mathbf{r}_j} \mathbf{G}(\mathbf{r}_j)$. 
Note that the resulting shift is small compared to the bare atomic transition frequency $|J_{\mathbf{k}}| \ll \omega_0$. Thus, the radiation is emitted at a frequency close to $\omega_0$ and the wavevector of the emitted photon needs to fulfill $|\boldsymbol{
\kappa}| \approx \omega_0/c=2\pi/\lambda_0$, which defines the edge of the light cone. For lattice spacings $d<\lambda_0/\sqrt{2}$, two types of momentum states emerge: first, those that lie inside the light cone $|\mathbf{k}|\leq 2\pi/\lambda_0$, which couple to the radiation field and have a nonzero decay rate, and second, those that lie outside the light cone $|\mathbf{k}|>2\pi/\lambda_0$ and only couple to evanescent waves of the electromagnetic field. 
As a result, these momentum states do not radiate into the electromagnetic far field and have suppressed decay rates. For an infinite lattice, perfect subradiance is achieved, $\Gamma_\mathbf{k}=0$, and the momentum states are completely dark \cite{Asenjo_2017_general,Janos_2017}.

\subsection{Coupling momentum states with a periodic spatial detuning profile}

In the case where all lattice atoms have the same detuning, the last term in Eq.~(\ref{eq: EquationMotionMomentumPrior}) is equal to $i \Delta (t) v_{\mathbf{k}}$ and every momentum state evolves independently. To couple different momentum states, one needs to break the symmetry of the lattice. This can be achieved by introducing a periodic spatial detuning profile. If the lattice is taken to be in the $xy$-plane, the most general periodic detuning profile with a periodicity of $N_x$ and $N_y$ atoms along each direction can be represented via the Fourier series

\begin{equation}
\label{eq: generalDetuning}
    \Delta_j = \sum_\mathbf{Q} \Delta_\mathbf{Q} e^{i \mathbf{Q} \mathbf{r}_j},
\end{equation}

\noindent where we have defined the two-dimensional quasimomenta $\mathbf{Q}=2\pi/d \left( q_x/N_x , q_y/N_y \right) $ such that $q_x \in \{0,N_x-1 \}$ and $q_y \in \{0,N_y-1 \}$. Note also that the detuning must be real at all lattice sites, which implies $\Delta_{(q_x,q_y)}= \Delta^*_{(N_x-q_x,N_y-q_y)}$.

Using the detuning in Eq.~(\ref{eq: generalDetuning}), the equations of motion in momentum space given by Eq.~(\ref{eq: EquationMotionMomentumPrior}) can be ultimately written as

\begin{equation}
\label{eq: eom_momentumspace_general}
    \frac{d v_{\mathbf{k}}}{dt}=  - i \left( J_{\mathbf{k}} - i \frac{\Gamma_\mathbf{k}}{2} \right) v_{\mathbf{k}} + i \sum_\mathbf{q} \Delta_\mathbf{q} v_{\mathbf{k}-\mathbf{Q}},
\end{equation}

\noindent such that the lattice state with momentum $\mathbf{k}=(k_x,k_y)$ is now coupled to states with momentum $\mathbf{k}-\mathbf{Q}= \left(k_x -\frac{2\pi}{d} \frac{q_x}{N_x} , k_y -\frac{2\pi}{d} \frac{q_y}{N_y} \right)$.
Having control over the periodicity of the pattern and the amplitude of each Fourier component allows to engineer a wide variety of couplings that can be used for various quantum applications discussed in subsequent sections. Note also that $q_x=q_y=0$ implies a constant detuning which does not break the symmetry of the system. As discussed above, this results in each momentum state evolving independently.

Finally, it is worth noting that this scheme can easily be generalized to atomic arrays of different geometries and dimensions.

\subsubsection{Checkerboard pattern}
\label{subsec: checkerboard}

To illustrate how different momentum states couple, we first consider a checkerboard detuning pattern such that nearest neighbors have detunings of opposite sign. The resulting non-Bravais square lattice with a two-atom basis is depicted in Fig.~\ref{fig: schematic}(a) and the exact form of the detuning profile can be expressed as $\Delta_j=\Delta e^{i \frac{\boldsymbol{\pi}}{d} \mathbf{r_j}} = \Delta (-1)^{n_j^x+n_j^y}$, where $\boldsymbol{\pi}=(\pi,\pi)$ and $\mathbf{n}_j=\mathbf{r}_j/d$ is an integer vector that represents the atomic positions normalized by the interparticle spacing. This pattern has a two-atom periodicity along both axes, contains a single Fourier component $\mathbf{Q}=\boldsymbol{\pi}/d$, and therefore couples the momentum states $|\mathbf{k}\rangle$ and $| \mathbf{k}+\boldsymbol{\pi}/d\rangle$.

The dynamics of the system are equivalent to those of a three-level system formed by the momentum states $|\mathbf{k}\rangle$ and $| \mathbf{k}+\boldsymbol{\pi}/d\rangle$ and by the lattice ground state $|g\rangle$, as depicted in Fig.~\ref{fig: schematic}(b). Each momentum state decays to the ground state with its corresponding decay rate and is shifted from the atomic frequency $\omega_0$ by its corresponding energy shift. Additionally, the coupling between both momentum states is given by the magnitude of the detuning $\Delta$. The corresponding equations of motion are   

\begin{align}
\label{eq: eom_checkerboard}
        \frac{d v_{\mathbf{k}}}{dt} & =  - i \left( J_{\mathbf{k}} - i \frac{\Gamma_\mathbf{k}}{2} \right) v_{\mathbf{k}} + i \Delta(t)  v_{\mathbf{k}+\boldsymbol{\pi}/d}, \nonumber \\
        \frac{d v_{\mathbf{k}+\boldsymbol{\pi}/d}}{dt} & =  - i \left( J_{\mathbf{k}+\boldsymbol{\pi}/d} - i \frac{\Gamma_{\mathbf{k}+\boldsymbol{\pi}/d}}{2} \right) v_{\mathbf{k}+\boldsymbol{\pi}/d} + i \Delta(t)  v_{\mathbf{k}}, \nonumber \\
        |g|^2 & =1-|v_{\mathbf{k}}|^2-|v_{\mathbf{k}+\boldsymbol{\pi}/d}|^2.
\end{align}

\noindent Note that population transfer from $|g\rangle$ to $|\mathbf{k}\rangle$ requires an external drive with in-plane wavevector $\mathbf{k}$.

To visualize the coupling between momentum states, it is useful to draw the lattice in reciprocal space. Figure~\ref{fig: schematic}(b) shows the first Brillouin zone for the nondetuned perfect lattice in black and for the detuned checkerboard lattice in blue. The detuning landscape couples the points along the two green, red, and orange dashed lines, which are translated by $\boldsymbol{\pi}/d$. For instance, the origin of the nondetuned Brillouin zone $\Gamma$ (a radiating state) is coupled to the corner $M$ (a dark state). Periodic detuning patterns therefore emerge as a natural technique to populate lattice dark states or, alternatively, to couple dark states to far-field radiation.

One can also understand the effect of the checkerboard detuning by analyzing the lattice band structure. Figures~\ref{fig: schematic}(c) and \ref{fig: schematic}(d) show the energy shift and decay rate (color scale) along the path $M' \rightarrow \Gamma \rightarrow X' \rightarrow M'$ of the detuned Brillouin zone. For zero detuning [Fig.~\ref{fig: schematic}(c)], only the momentum states inside the light cone, that is, close to the $\Gamma$ point, have a nonzero decay rate (yellow) and radiate. Additionally, the two bands are degenerate at the edges of the Brillouin zone, indicating that the detuned Brillouin zone does not capture the whole symmetry of the lattice and that the bands should be instead unfolded along the $M \rightarrow \Gamma \rightarrow X \rightarrow M$ path. For finite detuning [Fig.~\ref{fig: schematic}(d)], the lattice loses the full symmetry and a band gap opens. 
The momentum states of the nondetuned lattice $v_\mathbf{k}$ are no longer eigenstates of the system and the two bands are now admixtures of $v_\mathbf{k}$ and $v_{\mathbf{k}+\boldsymbol{\pi}/d}$. Near the $\Gamma$ point, this results in both bands acquiring a finite width, such that initial dark states close to $M$ are forced to radiate. For small enough lattice constants, a region appears where dark momentum states are coupled with one another and remain subradiant despite the detuning pattern.

\subsection{Emitted electric field}

The atomic dynamics determine the electric field at any given point and time through Eq.~(\ref{eq: electricfield_operator}). When Fourier transformed over the in-plane spatial components $x$ and $y$, the field can be written as

\begin{equation}
\label{eq: field_finite}
    \mathbf{E}(\boldsymbol{\kappa}_{||},z,\omega)=\mu_0 k_0^2 \mathbf{G}(\boldsymbol{\kappa}_{||},z,\omega_0) \mathbf{d} \sum_j e_j(\omega-\omega_0) e^{-i \boldsymbol{\kappa}_{||}\mathbf{r}_j},
\end{equation}

\noindent where $\boldsymbol{\kappa}_{||}=(\kappa_x,\kappa_y)$ is a two-dimensional momentum vector, $\mathbf{G}(\boldsymbol{\kappa}_{||},z,\omega_0)$ is the momentum-space Green's function and $e_j(\omega-\omega_0)=\int_0^\infty dt e^{i (\omega-\omega_0) t} e_j(t)$ is the shifted Laplace transform of the atomic amplitudes. Note that we have again assumed that the atomic response is narrow around its resonant value $\omega_0$. For an infinite lattice, the summation over atomic sites corresponds to the definition of the lattice momentum states and Eq.~(\ref{eq: field_finite}) takes the simple form

\begin{equation}
\label{eq: field_infinite}
    \mathbf{E}(\boldsymbol{\kappa}_{||},z,\omega)=\mu_0 k_0^2 \mathbf{G}(\boldsymbol{\kappa}_{||},z,\omega_0) \mathbf{d} v_{\boldsymbol{\kappa}_{||}}(\omega-\omega_0).
\end{equation}

\noindent That is, a lattice state with a given quasimomentum $\mathbf{k}$ couples only to modes of the electromagnetic field with that same in-plane momentum $\boldsymbol{\kappa}_{||}=\mathbf{k}$. Because the three-dimensional momentum $\kappa$ fulfills $|\boldsymbol{\kappa}| \leq 2\pi/\lambda_0$, only states inside the lightcone can couple to the electromagnetic field and radiate. Note also that the excitation will be identically emitted to each side of the array such that the amplitudes at mirrored directions $({\boldsymbol{\kappa}_{||}}, \pm \kappa_z)$ are equal.

In order to characterize the electric field or photon emitted by the array, we will use three figures of merit in what follows. The first is the magnitude of the electric field as a function of frequency for a fixed propagation direction, which is given by the collective frequencies at which the radiating momentum states evolve. The second is the magnitude of the electric field for different propagation directions at a fixed frequency, which primarily depends on the amplitude of each radiating momentum state. The third is the overlap between the field generated by the atomic array and the electric field of a mode of interest, which we refer to as the detection mode. 
This overlap only depends on the amplitude of the detection mode at the atomic positions \cite{Manzoni_2018}. For a monochromatic Gaussian beam of waist $\rho$ that propagates perpendicular to the atomic array and whose focal plane coincides with the position of the lattice, the amount of excitation in the detection mode as a function of time can be written as \cite{Manzoni_2018}

\begin{equation}
\label{eq: efficiency}
        n(t)=\frac{3}{4 \pi^2 \rho^2}  \int_0^t d\tau \left | \sum_{j} \mathbf{d} \boldsymbol{\epsilon}_G^* e^{-\mathbf{r}^2/\rho^2} e_j(\tau) \right |^2,
\end{equation}

\noindent where $\boldsymbol{\epsilon}_G$ is the polarization of the Gaussian mode. The long-time limit $\eta= \lim_{t\rightarrow \infty} n(t)$ gives the efficiency of the emission process, that is, the fraction of the photon emitted into the desired mode.

\section{\label{sec: memory} Single-photon storage and retrieval}

In the previous section, we have presented a general method to couple different momentum states with one another. In particular, such spatial detuning patterns allow populating lattice dark states, that is, momentum states outside the light cone that in general do not couple to the electromagnetic far field. Using the checkerboard pattern presented in Sec.~\ref{subsec: checkerboard}, an incident beam with in-plane momentum $\mathbf{k}$ 
will not only populate the radiating lattice state with that same quasimomentum $\mathbf{k}$, but also the dark state $\mathbf{k}+\boldsymbol{\pi}/d$. For example, a low-intensity, classical Gaussian beam with zero in-plane momentum will populate a Gaussian distribution of momentum states centered at the corner of the Brillouin zone $M$ (see Appendix~\ref{app: classicalDrive}). Upon turning off the spatial detuning pattern, the long-lived dark states will remain in the lattice.

This selective coupling to lattice dark states also enables the storage of traveling, single photons in the atomic array. In recent decades, disordered, three-dimensional atomic clouds \cite{Gorshkov_PRA_2007,Gorshkov_PRL_2007}, as well as two-dimensional ordered atomic arrays \cite{Manzoni_2018} have been proposed as light-matter interfaces for photon storage. While atomic clouds require large optical depths to achieve near-unity storage and retrieval fidelity, strong interactions and spatial interference can be exploited in atomic arrays to obtain similar efficiencies with a very moderate amount of atoms ($\sim 100$ atoms). Both platforms use a similar storage scheme, wherein electromagnetically induced transparency enables the storage of the photon in a metastable, third atomic level. We now present a different storage mechanism that requires only two atomic levels and that uses lattice dark states to convert the incoming photon in a long-lived collective atomic excitation.

For simplicity, let us first study the retrieval process. Consider a collective atomic excitation at the corner of the Brillouin zone $M$. Once the checkerboard detuning pattern is turned on, the excitation will couple to the origin of the Brillouin zone $\Gamma$ and radiate. Due to the symmetry of the array, the photon will be emitted on both sides of the lattice and needs to be recombined \emph{a posteriori}. The temporal shape of the photon will be determined by the precise form of $\Delta(t)$, and the efficiency of the process is obtained by comparing the outgoing photon with a mode of interest (i.e., a Gaussian mode). Conversely, photon storage can be simply understood as the time-reversed operation of retrieval \cite{Gorshkov_PRA_2007,Gorshkov_PRL_2007}. That is, sending the photon back in a time-reversed way and applying the time-reversed detuning sequence ensures the storage of the photon. It follows that the efficiencies for storage and retrieval are identical. 

\begin{figure}[b]
\includegraphics[width=\linewidth]{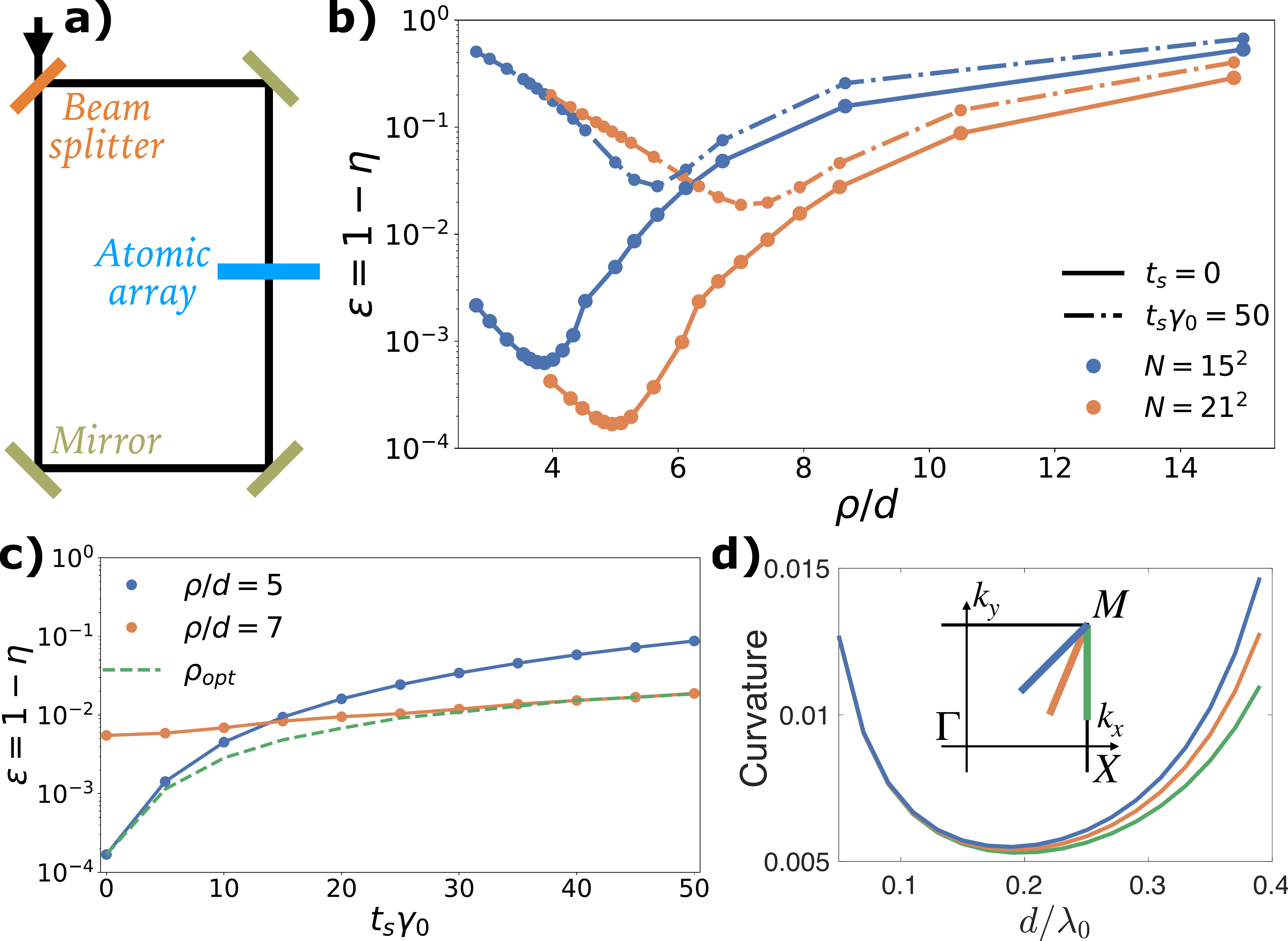}
\caption{\label{fig: memory} \textbf{Experimental setup and efficiency of photon retrieval.} (a) Setup to store a photon in the two-dimensional atomic array. The incoming photon is split by a beam splitter and impinges the lattice from both sides. Conversely, the retrieved or outgoing photon is equally emitted on both sides of the array and needs to be recombined. (b) Retrieval error of a single photon in Gaussian detection modes of waist $\rho$ for lattices with spacing $d=0.3\lambda_0$ and different total number of atoms, $N$. In all figures, we consider the atoms to be circularly polarized, $\wp=(1,1j,0)/\sqrt{2}$. The solid lines represent the error if the Gaussian photon is retrieved immediately after being stored ($t_s=0$). The dashed lines show the error if a finite amount of time $t_s\gamma_0=50$ goes by before retrieving the photon. (c) Retrieval error as a function of storage time for waists $\rho/d=5$ (blue) and $\rho/d=7$ (orange), as well as for the optimal waist $\rho_{opt}$ for any given storage time $t_s$ (green). A $21 \times 21$ lattice is considered. (d) Curvature of the band around the Brillouin zone corner $M$ as a function of lattice spacing along the directions shown in the inset. The band curvature determines the rate at which the different momentum components dephase. Lattices with larger curvatures and stored excitations with larger momentum width dephase faster.}
\end{figure}

This implies that a single photon can only be perfectly stored in the two-dimensional lattice if it impinges from both sides \cite{Manzoni_2018,setup_bothsides}. The incoming photon therefore needs to be split using the experimental setup depicted in Fig.~\ref{fig: memory}(a). Moreover, given an incoming photon, one needs to find the detuning sequence $\Delta(t)$ capable of storing it or, equivalently, capable of producing that same temporal shape during the retrieval process. The resulting optimization problem is treated in Sec.~\ref{subsec: temporalprofile}.

In what follows, we evaluate the retrieval efficiencies that can be obtained for various lattice sizes, as well as the inherent limitations of photon memories with lattice dark states.

\subsection{Retrieval efficiency}

To evaluate the retrieval efficiency of the array, we compute the overlap of the emitted photon with various detection modes of Gaussian shape. The amount of excitation emitted in the desired channel as a function of time is given by Eq.~(\ref{eq: efficiency}) and its limit at long times gives the efficiency $\eta$ of the process. We consider square arrays of different atom number $N$, as well as Gaussian beams of different waists $\rho$. 
The corresponding initial dark states are obtained by illuminating a deexcited lattice with a low-intensity classical Gaussian beam of the desired waist.
As shown in Appendix~\ref{app: classicalDrive}, applying the checkerboard detuning pattern results in a Gaussian-like superposition of dark momentum states around the Brillouin zone corner $M$.

Figure~\ref{fig: memory}(b) shows the retrieval error $\epsilon=1-\eta$ of such dark states as a function of beam waist $\rho$ for different array sizes. We retrieve the same bounds as in Ref.~\cite{Manzoni_2018}, where the excitation is stored in a third, metastable level. 
For beam waists larger or comparable to the array size, the error corresponds to the amount of energy that propagates outside the array and does not interact with the lattice. At low beam waists, the Gaussian beam contains wavevectors with many different propagation directions. Different angles of incidence have maximum reflectance at different detunings \cite{Ephi_2017}, which results in a reduced overall reflectance of the photon and a subsequent increase of the retrieval error. There exists an intermediate region where both error sources are simultaneously reduced and minimal errors are obtained. Larger lattices reduce the amount of energy lost outside of its boundaries and enable the storage of Gaussian beams with larger waists, and therefore lower momentum spreads. As a result, larger lattices attain lower errors and the optimal waist scales with the array size. Note that one can achieve storage and retrieval errors of the order $\epsilon \sim 10^{-4}$ for $21 \times 21$ lattices, that is, arrays of 441 atoms. Provided that the spacing $d$ is small enough such that the lattice presents subradiant modes, the exact value of the efficiency only depends on the ratio $\rho/d$ and the size $N$ of the array.

While the efficiency is mostly unaffected by the specific detuning $\Delta(t)$ applied during the retrieval process, it strongly depends on the spatial shape of the incoming photon. A photon with a finite waist is stored in a superposition of dark momentum states, each of them having a different energy shift $J_\mathbf{k}$ from the bare atomic frequency $\omega_0$. Therefore, once the photon is mapped to a collective atomic excitation and the spatial detuning is turned off, each momentum component freely evolves at a different frequency $v_{\mathbf{k}}(t)=e^{-iJ_{\mathbf{k}} t }v_{\mathbf{k}}(t=0)$. As a result, different momentum components acquire different relative phases while the excitation remains stored in the lattice and the system dephases. If the excitation is retrieved immediately after being stored, all momentum components have the original phases and the photon is emitted almost perfectly into its original mode [solid lines in Fig.~\ref{fig: memory}(b)]. If one waits a finite amount of time $t_s$ before starting the retrieval process, this is no longer the case and the spatial profile of the emitted photon depends on the exact shape of the bandstructure at the region where the excitation was stored. Around the $M$ point, the dispersion is parabolic and the resulting field no longer has the perfect Gaussian profile of the stored photon. Consequently, the efficiency of the retrieval process decreases with $t_s$. As shown by Fig.~\ref{fig: memory}(c) and the dash-dotted curves in Fig.~\ref{fig: memory}(b), this decay in efficiency is lower for photons with a large waist, as they excite a narrower distribution of dark momentum states. The minimum error for a $21 \times 21$ lattice with $d=0.3\lambda_0$ as a function of storage time is given by the green curve in Fig.~\ref{fig: memory}(c) and reaches $\sim 2\%$ for $t_s\gamma_0=50$, a value that can be significantly reduced by increasing the size of the lattice. Note also that the dephasing rate in square lattices is given by the curvature of the band at the $M$ point, which is minimal for spacings $d \sim 0.2\lambda_0$ [see Fig.~\ref{fig: memory}(d)]. Alternatively, the dephasing can be minimized by considering different lattice geometries and exploiting the properties of their bandstructures. For example, storing the photon in momentum states lying on flat bands would eliminate dephasing and allow to retrieve the photon after large storage times with almost unit fidelity.

\section{\label{sec: shaping} Shaping a single photon}

One of the fundamental goals of quantum optics is to control the state of nonclassical light. In this section, we discuss how two-dimensional atomic lattices can be used to modify the properties of single photons by choosing different temporal detuning sequences and spatial detuning patterns for the storage and retrieval processes. More concretely, we show how to create photons with arbitrary temporal shapes, as well as how to produce states of light relevant in multidimensional quantum information science \cite{multidimensionalphotoncontrol_1,multidimensionalphotoncontrol_2,multidimensionalphotoncontrol_3,controlsinglephoton_1,controlsinglephoton_2,controlsinglephoton_3,controlsinglephoton_4} such as photons that are in a coherent superposition of different frequencies or directions.

\subsection{Temporal shape}
\label{subsec: temporalprofile}

For a given spatial detuning pattern, the amplitude of the detuning determines the strength of the coupling between momentum states and the dynamics of the retrieval process.
As a result, a photon with a given temporal shape 
can only be achieved by some detuning sequence $\Delta(t)$. As discussed in Sec.~\ref{sec: memory}, this knowledge is necessary to properly store an incoming photon and to later retrieve it with the same temporal shape. Additionally, one can also manipulate the temporal shape of the outgoing photon at-will by modifying $\Delta(t)$.

For simplicity, we will first consider an infinite lattice with an excitation initially stored in a dark state with momentum $\mathbf{k}_d$. Applying a checkerboard detuning pattern, the dark state will couple to the state $\mathbf{k}_r=\mathbf{k}_d+\boldsymbol{\pi}/d$, which will radiate if it is located within the light cone. The equations of motion for the three-level system containing the momentum states $\mathbf{k}_d$ and $\mathbf{k}_r$ and the ground state $|g\rangle$ are given by Eq.~(\ref{eq: eom_checkerboard}). To simplify the notation, we will denote the amplitude in the dark state $v_{\mathbf{k}_d} \equiv v_d$ and the amplitude in the radiating state $v_{\mathbf{k}_r} = v_{\mathbf{k}_d+\boldsymbol{\pi}/d} \equiv v_r$. Similarly, the energy shifts and decay rates of both states will be labeled as $J_d$, $J_r$, and $\Gamma_r$ (note that $\Gamma_d=0$).

The amount of excitation in the radiation field is equal to the amount of excitation that has left the system. If the photon is retrieved into the detection mode with high efficiency, the derivative of the photon number in the detection mode $dn/dt$ is given by

\begin{equation}
\label{eq: photonshape_excitedpopulation}
        \frac{dn}{dt} \approx \frac{d|g|^2}{dt}=-2\mathbf{Re} \{ \dot{v}_d v_d^* +\dot{v}_r v_r^*\} = \Gamma_{r} |v_r|^2.
\end{equation}

\noindent That is, the desired temporal shape of the photon fixes the population of the radiating state during the retrieval process and can therefore be related to the detuning sequence $\Delta(t)$ needed to produce that photon. For that, we discretize the time evolution in small steps of time $\delta t$. At each time step $k$, the amplitudes in the momentum states are

\begin{align}
        v_d^{(k)}&=v_d^{(k-1)}+ \delta t \left( -i J v_d^{(k-1)} + i \Delta^{(k-1)} v_r^{(k-1)} \right), \nonumber\\
        v_r^{(k)}&=v_r^{(k-1)}+ \delta t \left( i \Delta^{(k-1)} v_d^{(k-1)} - \Gamma_{r} v_r^{(k-1)}/2\right),
\end{align}

\noindent where we have defined $J=J_d-J_r$. Plugging these expressions in the discretized version of Eq.~(\ref{eq: photonshape_excitedpopulation}) results in a second-order equation for the detuning at step $k-1$ with solution 

\begin{align}
\label{eq: numericalmethod}
    \Delta^{(k-1)}&= \left(-b \pm \sqrt{b^2-4ac}\right)/2a, \nonumber \\
    a&=\delta t^2 |v_d^{(k-1)}|^2, \nonumber\\
    b&=2 \delta t  \left( 1- \Gamma_{r} \delta t/2\right)  \mathbf{Im} \{ v_r^{(k-1)} v_d^{(k-1)*} \}, \nonumber\\
    c&=- \frac{dn/dt |_k}{\Gamma_{r}} + |v_r^{(k-1)}|^2 \left( 1- \Gamma_{r} \delta t/2\right)^2,
\end{align}

\noindent with initial conditions $v_d(t=0)=1$ and $v_r(t=0)=0$. For this result to be valid in the case of a finite lattice and a stored state with a finite momentum width, one requires the decay rate $\Gamma_r$ and the difference in energy shifts $J$ to be approximately constant over the distribution of momenta contained in the initial dark state. The band structure in Fig.~(\ref{fig: schematic})(c) shows that this is the case for an excitation initially stored at the corner of the Brillouin zone $M$ and that couples to the origin $\Gamma$.

\begin{figure}[b]
\includegraphics[scale=0.28]{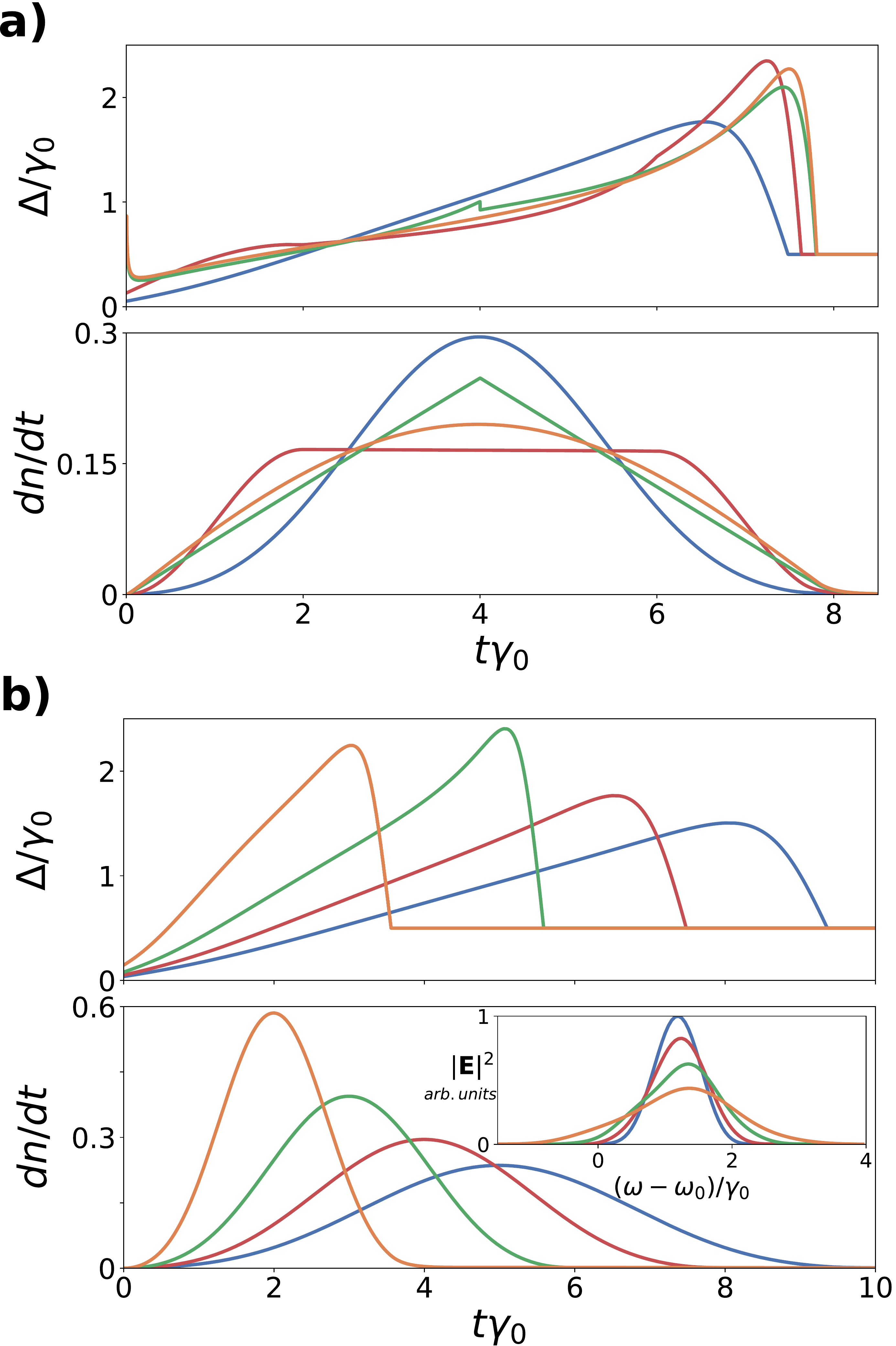}
\caption{\label{fig: temporal} \textbf{Temporal shape of the emitted photon.} (a) Optimized temporal detuning sequences $\Delta(t)$ for an infinite lattice and emitted photons of the following shapes: Blackman window (blue), Tukey window (red), triangular window (green), and sinusoidal window (orange). Lower plot: Population emitted in a Gaussian mode of waist $\rho=1.2\lambda_0$ per unit of time $dn/dt$ for the different detuning sequences and for a $21 \times 21$ lattice with lattice spacing $d=0.2\lambda_0$. (b) Detuning sequences optimized for a finite lattice and resulting temporal shape of the emitted photon for Blackman windows of different durations. Again, a $21 \times 21$ lattice with lattice spacing $d=0.2\lambda_0$ is used. The inset shows the spectrum of the emitted electric field. The resulting photon has only one central frequency and its width increases as its temporal duration decreases.
}
\end{figure}

Using the $M$ and $\Gamma$ points as a reference, we numerically obtain in Fig.~\ref{fig: temporal}(a) the detuning sequences for photons of the following shapes: Blackman window (blue), Tukey window (red), triangular window (green), and sinusoidal window (orange). We then consider a $21 \times 21$ square lattice of spacing $d=0.2\lambda_0$ with a stored Gaussian beam of waist $\rho=1.2 \lambda_0$. Applying the optimized sequences to this finite lattice results in the temporal photon shapes $dn/dt$ in the lower plot of Fig.~\ref{fig: temporal}(a), which perfectly resemble the target shapes.
Note that the exponential nature of the atomic decay makes it challenging to obtain a temporal shape that abruptly finishes at a given finite time $t_{end}$. For such shapes, where $dn/dt(t \leq t_{end}) \neq 0$ and $dn/dt(t>t_{end}) = 0$, the numerical method in Eq.~(\ref{eq: numericalmethod}) yields detuning strengths that significantly grow close to $t_{end}$. Here, we replace these growing endings by constant detuning plateaus at $\Delta=0.5\gamma_0$. While the effect is negligible for temporal shapes with a smooth tail such that $d^2n/dt^2(t = t_{end})$ is continuous (i.e., Blackman and Tukey windows), this is no longer true for shapes that exhibit sharp endings with a discontinuous $d^2 n/dt^2(t = t_{end})$ . That is, the exponential nature of the atomic array forces the triangular and sinusoidal photons to also have a smooth tail. Additionally, the discontinuity of $d^2 n/dt^2$ at the peak of the triangular photon requires a jump in $\Delta(t)$.

Figure~\ref{fig: temporal}b shows the detuning and photon shape for Blackman windows (similar to a confined Gaussian window) of different duration. While the emission is centered around a single frequency in all cases, shorter temporal profiles require larger magnitudes of the detuning and their spectrum consequently has a larger linewidth [see inset  Fig.~\ref{fig: temporal}(b)].

Note that all photons generated in Fig.~\ref{fig: temporal} have the same spatial profile, which is uniquely defined by the spatial detuning pattern (in this case a checkerboard pattern) and the storage time, during which the relative phases between momentum states are modified.

\subsection{Frequency profile}

While applying time-dependent detuning sequences such that at all times $\Delta(t) < \Gamma_r$ produces photons centered at a single frequency, this is no longer the case for sufficiently large magnitudes of the detuning.
From a band structure perspective (see Fig.~\ref{fig: schematic}), increasing the magnitude of the detuning also increases the gap between the bands. For large enough detunings, the frequency difference between the bands will be larger than their widths or decay rates and one would expect the emission to happen at two different frequencies.

To analyze this phenomenon, let us consider that a dark state and a radiating state of the infinite lattice are coupled via a checkerboard detuning pattern of constant strength $\Delta$. Then, Eq.~(\ref{eq: eom_checkerboard}) can be solved analytically and the amplitude in the radiating state $v_r$ can be written as

\begin{equation}
    v_r(t)=\frac{\Delta}{2\sqrt{\Delta^2+G^2}} \left(e^{i \omega_+ t} -e^{i \omega_- t} \right),
\end{equation}

\noindent where we have defined $G=\frac{J_d-J_r+i\Gamma_r/2}{2}$ and $\omega_{\pm}=\frac{-J_d-J_r+i\Gamma_r/2}{2} \pm  \sqrt{\Delta^2+G^2}$. Note that the real and imaginary parts of $\omega_\pm$ correspond to the energy shifts and decay rates of the two bands of the checkerboard lattice. 

The photon will be emitted at the in-plane momentum of the radiating state $\mathbf{k}_r$ and its frequency at late times can be obtained from Eq.~(\ref{eq: field_infinite}):

\begin{equation}
\label{eq: freq_profile}
    \mathbf{E}_{\boldsymbol{\kappa}_{||}=\mathbf{k}_r} (\omega)  \propto   \frac{1}{\omega - \omega_0 + \omega_+}   -  \frac{1}{\omega - \omega_0 + \omega_-} .
\end{equation}

\noindent It is now clear that the spectrum will have two Lorentzian peaks. Their centers are given by the real part of $\omega_{\pm}$, whereas the magnitude and width of the peaks are determined by the imaginary part of the $\omega_{\pm}$.

\begin{figure}[b]
\includegraphics[width=\linewidth]{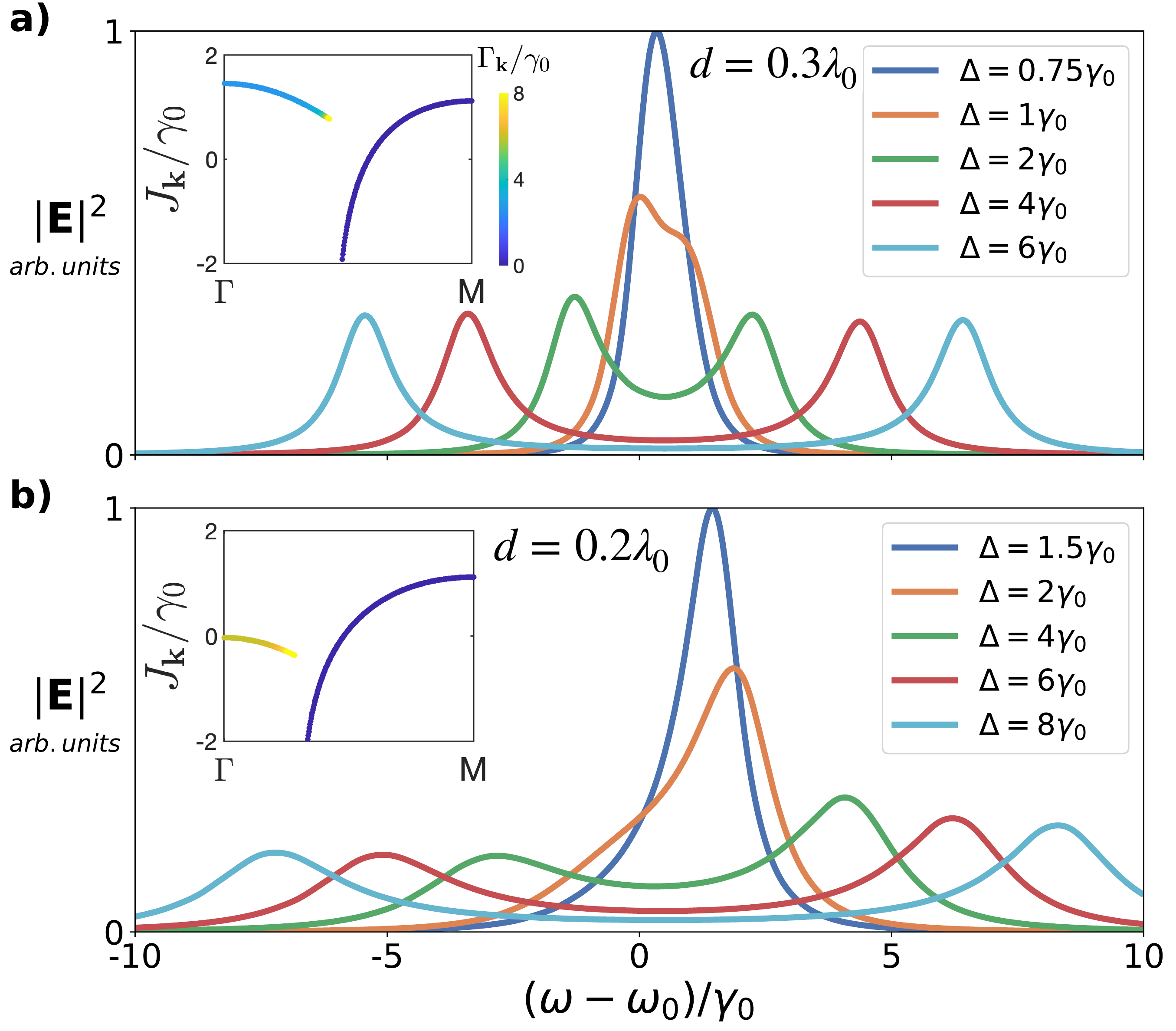}
\caption{\label{fig: frequency} \textbf{Frequency profile of the emitted photon.} Frequency profile of the emitted electric field for different detuning strengths and for spacings (a) $d=0.3\lambda_0$ and (b) $d=0.2\lambda_0$. As the detuning is increased, the central frequency peaks splits into two. The insets show the band structure of the lattice with zero detuning. The sign of $J_\Gamma-J_M$ determines whether the most prominent peak is found at frequencies larger or lower than the natural transition frequency $\omega_0$. The same color scale is used for both insets.}
\end{figure}

Figures~\ref{fig: frequency}(a) and \ref{fig: frequency}(b) show the frequency profile of the photon emitted by a $21 \times 21$ lattice with spacing $d=0.3\lambda_0$ and $d=0.2\lambda_0$, respectively. As expected, only one frequency peak is observed for low detunings such that $\Delta \ll \Gamma_r$. If the detuning is increased, the frequency difference between both bands becomes larger than their decay rates and two asymmetric peaks emerge, indicating that the outgoing photon is in a superposition of two colors. For a very large detuning, the decay rates of both bands tend to $\Gamma_r/2$ and the probability to emit at each frequency is the same. In this limit, we can think of the atomic array as being formed by two independent square lattices of spacing $\sqrt{2}d$. Their resonance frequencies are shifted by $2\Delta$ and their decay rates are identical and equal to $\Gamma_r/2$. Note that arrays with a smaller spacing have a larger decay rate and therefore require a larger detuning before emission in two colors is observed. Also, the most prominent peak corresponds to the band with the original dark state. Depending on the spacing, the dark state can have a larger or lower energy shift than the radiating state and the predominant peak will correspondingly appear at a frequency larger or lower than $\omega_0$.

Additional control of the frequency profile can be achieved by adding a periodic time-dependent detuning \cite{Silveri_2017_modulation,freqmodulation_Lukin} that is equal at all lattice sites. In Appendix~\ref{app: frequency modulation}, we show that it results in 
modulation-shifted sidebands and that the photon is emitted in a coherent superposition of many frequencies.

\subsection{Direction of emitted photon}
\label{sec: direction}

\begin{figure}[b]
\includegraphics[width=\linewidth]{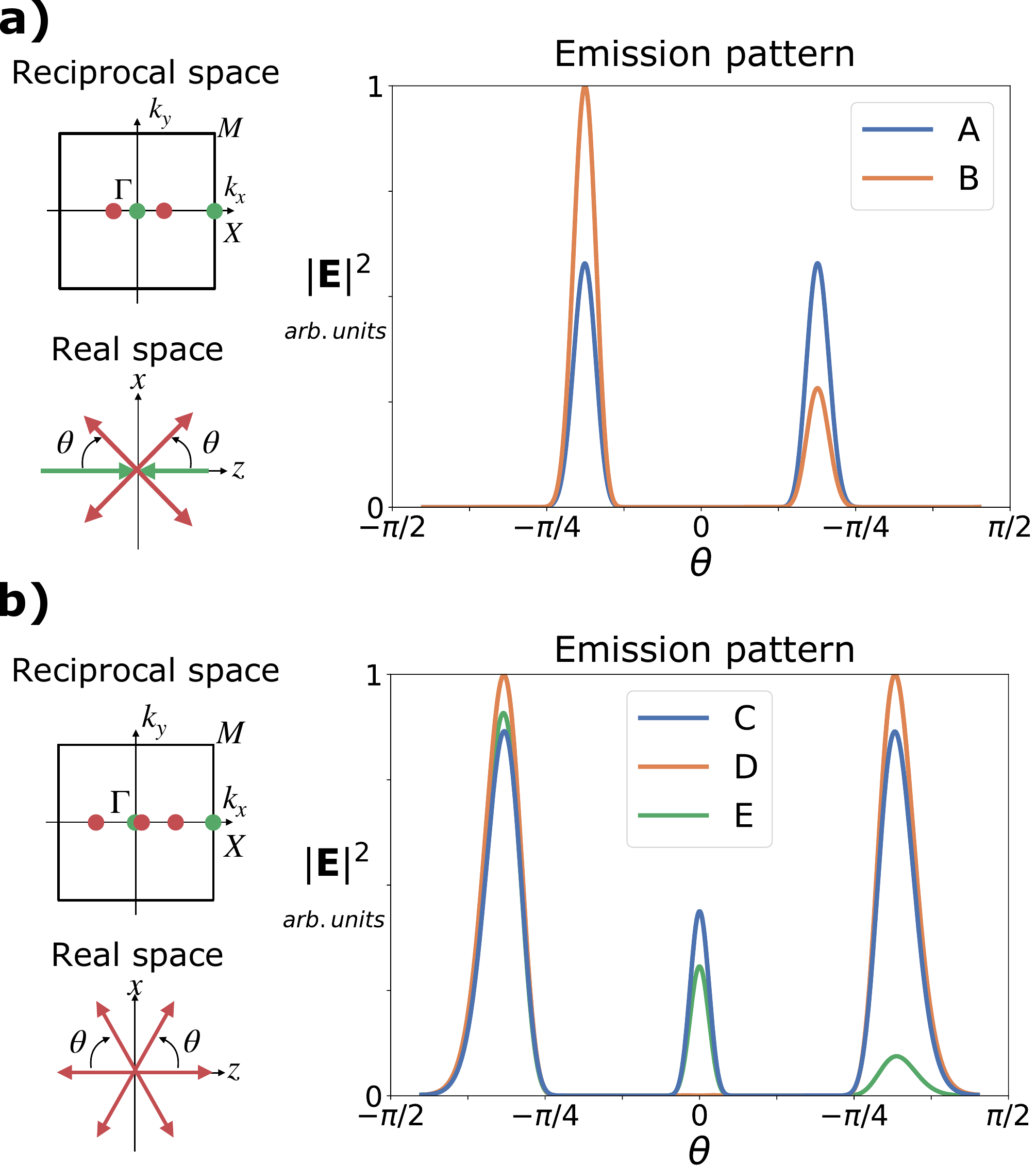}
\caption{\label{fig: directions} \textbf{Direction of the emitted photon.} (a) The photon is initially stored at the $X$ point by coupling it to the origin of the Brillouin zone $\Gamma$ (green dots). Applying spatial detunings with a periodicity of three atoms along the $x$ direction, the $X$ point is coupled to the two red dots with momentum $\mathbf{k}=(\pm \pi/3d,0)$. In real space, the photon to be stored impinges the array in the perpendicular direction (green arrows), whereas the excitation is retrieved in a superposition of the four red arrows which form an angle $\sin \theta \approx \lambda_0/6d$ with the perpendicular axis. We compute the magnitude of the electric field along the directions $\boldsymbol{\kappa}_\theta=2\pi/\lambda_0 (\sin \theta,0,\pm \cos \theta)$ contained in the $xz$ plane for two different detunings: $\Delta_A/\gamma_0=e^{i\mathbf{Q_+}\mathbf{r}_j}/2+e^{i\mathbf{Q_-}\mathbf{r}_j}/2$ (blue) and $\Delta_B/\gamma_0= -i e^{i\mathbf{Q_+}\mathbf{r}_j}/2+i e^{i\mathbf{Q_-}\mathbf{r}_j}/2$ (orange), where $\mathbf{Q}_{\pm}=(\pm 2\pi/3d,0)$. (b) Same for detuning patterns with a periodicity of four atoms.
The $X$ point is in this case coupled to the origin $\Gamma$ and the states $\mathbf{k}=(\pm \pi/2d,0)$ and the photon is emitted in a superposition of six directions, two perpendicular to the array and four which form an angle $\sin \theta \approx \lambda_0/4d$ with the normal axis. The most general detuning pattern is $\Delta_j= \beta e^{i \tilde{\mathbf{Q}}_+ \mathbf{r}_j} + \beta^* e^{i \tilde{\mathbf{Q}}_- \mathbf{r}_j} + \delta e^{i \mathbf{Q}_\pi \mathbf{r}_j}$, where $\tilde{\mathbf{Q}}_{\pm}=(\pm \pi/2d,0)$ and $\mathbf{Q}_{\pi}=( \pi/d,0)$. Three detuning patterns are considered: pattern C with $\beta_C=0.75 \gamma_0$ and $\delta_C=0$, pattern D with $\beta_D=0.75 e^{i\pi/4}\gamma_0$ and $\delta_D=0$, and pattern E with $\beta_E=0.5 e^{i\pi/4}\gamma_0$ and $\delta_E=0.7\gamma_0$. In all cases, $41 \times 41$ lattices with spacing $d=0.3\lambda_0$ are used and the stored Gaussian photon has a waist of $\rho=12d$. }
\end{figure}

While modifying the magnitude of the detuning over time gives control over the time-frequency degrees of freedom of the emitted photon, it cannot modify its direction or spatial profile $\mathbf{E}(\mathbf{r})$. This can only be achieved by using different spatial detuning patterns for the storage and retrieval processes. 
In this section, we store a perpendicularly incoming Gaussian photon at the $X$ point of the Brillouin zone [green circles in Fig.~\ref{fig: directions}(a)] by applying a pattern with a single Fourier component $\mathbf{Q}=(\pi/d,0)$, which creates a two-atom periodicity along the $x$ axis given by $\Delta_{S}=\Delta (-1)^{n_j^x}$. Using retrieval patterns of a higher periodicity, we show that the outgoing photon can be produced in a coherent superposition of different directions.

In general, a detuning pattern with $n$ atoms per unit cell will couple $n$ momentum states with each other. That is, an initial dark state will be coupled to a set of $n-1$ momentum states. Provided that all of them lie inside the light cone, the photon will in general be emitted in a coherent superposition of $n-1$ directions (on each side of the lattice) given by the in-plane momenta of the radiating momentum states. The percentage of the photon emitted in each direction can be controlled by properly engineering the shape of the detuning pattern.

The first nontrivial detuning corresponds to a periodicity of three atoms, for example, along the $x$ axis. The three possible Fourier components are $\mathbf{Q}_0=\mathbf{0}$ and $\mathbf{Q}_{\pm}=(\pm 2\pi/3d,0)$. The dark state with momentum $\mathbf{k}=(\pi/d,0)$ will therefore be coupled to the momentum states $\mathbf{k}=(\pm \pi/3d,0)$ illustrated by the red circles of the Brillouin zone in Fig.~\ref{fig: directions}(a). For $d > \lambda_0/6$, these momentum states lie within the light cone and the photon will be emitted along the directions $\boldsymbol{\kappa} \approx 2\pi/\lambda_0(\pm \sin \theta,0,\pm \cos \theta)$, which correspond to the four red arrows (two on each side of the array) in the real-space sketch of Fig.~\ref{fig: directions}(a). They form an angle of $\sin \theta = \lambda_0/6d$ with the axis perpendicular to the array. To simplify nomenclature, we will label the different momentum states according to their normalized momentum $k_x d$ along the $x$-axis, such that $v_\pi$ is the dark state and $v_{\pm \pi/3}$ the radiating states. For a general detuning $\Delta_j= \alpha e^{i \mathbf{Q}_0 \mathbf{r}_j} + \beta e^{i \mathbf{Q}_+ \mathbf{r}_j} + \beta^* e^{i \mathbf{Q}_- \mathbf{r}_j} =\alpha + \beta e^{i 2\pi n_j^x/3 }+\beta^* e^{-i 2\pi n_j^x/3 }$ such that $\alpha \in \mathbb{R}$ and $\beta \in \mathbb{C}$, the equations of motion are

\begin{align}
\label{eq: direction_3atomperiod}
        \frac{dv_\pi}{dt}&=-i \left( J_\pi - \alpha \right) v_\pi + i \operatorname{Re}( \beta ) v_+ - \operatorname{Im} ( \beta ) v_-, \nonumber \\
        \frac{dv_+}{dt}&=-i \left( J_{\pi/3} -\alpha - \operatorname{Re}(\beta) -i \frac{\Gamma_{\pi/3}}{2} \right) v_+ \nonumber \\
        & + 2i \operatorname{Re}( \beta ) v_\pi + \operatorname{Im} ( \beta ) v_-, \nonumber \\
        \frac{dv_-}{dt}&=-i \left( J_{\pi/3} -\alpha + \operatorname{Re}(\beta) -i \frac{\Gamma_{\pi/3}}{2} \right) v_+ \nonumber \\
        & + 2 \operatorname{Im}( \beta ) v_\pi - \operatorname{Im} ( \beta ) v_+,
\end{align}

\noindent where we have introduced the rotated momentum basis $v_{\pm} = v_{\pi/3} \pm v_{-\pi/3}$. From Eqs.~(\ref{eq: direction_3atomperiod}), it is clear that the amount of excitation emitted into each direction depends on the parameter $\beta$. For $\beta \in \mathbb{R}$, $v_\pi$ couples only to $v_+$. If the excitation is initially stored in the dark state, then $v_-(t)=0$ at all times and the amplitudes in the radiating states are always the same, $v_{\pi/3}= v_{-\pi/3}$. The resulting photon is therefore emitted in an equal superposition of the four directions, as shown by the emission pattern A in Fig.~\ref{fig: directions}(a). For a general complex $\beta$, all three states are coupled with one another and it is no longer true that $|v_{\pi/3}|= |v_{-\pi/3}|$. The probability to emit the photon at an angle $\theta$ and $-\theta$ will therefore be different [see pattern B in Fig.~\ref{fig: directions}(a)] and the electric field will have an asymmetric spatial profile. Note that the relative amplitudes between both directions can be controlled by adjusting the magnitude of the detuning $\beta$.

To obtain simultaneous emission perpendicular to the array and along two oblique directions, one needs to introduce a detuning pattern with a periodicity of four atoms along the $x$ direction. Now, the Fourier components $\mathbf{Q}_0=0$, $\tilde{\mathbf{Q}}_\pm=(\pm \pi/2d,0)$, and $\mathbf{Q}_\pi=(\pm \pi/d,0)$ couple the stored dark momentum state $\mathbf{k}=(\pi/d,0)$ with the states $\mathbf{k}=(\pm \pi/2d,0)$ and $\mathbf{k}=\mathbf{0}$, which correspond to the red  circles in Fig.~\ref{fig: directions}(b). For $d > \lambda_0/4$, these momentum states lie within the light cone and the photon will be emitted either at the perpendicular direction $\theta=0$ or at an angle $\pm \theta$ such that $\sin \theta = \lambda_0/4d$. Again, different types of interference effects between momentum states can be engineered to obtain arbitrary electric fields. For a general detuning pattern containing only the Fourier components $\tilde{\mathbf{Q}}_\pm$, the photon will be emitted along the oblique directions $\pm \theta$ (with equal amplitude) and along the perpendicular direction, as shown by trace C in Fig.~\ref{fig: directions}(b). For certain values of the detuning, however, the momentum states $\mathbf{k}=(\pm \pi/2d,0)$ destructively interfere to suppress the amplitude in state $\mathbf{k}=\mathbf{0}$. The resulting electric field, given by trace D in Fig.~\ref{fig: directions}(b), is zero along the perpendicular direction. Finally, one can break the symmetry between the oblique directions $\pm \theta$ by introducing a component $\mathbf{Q}_\pi$ in the spatial detuning pattern. Then, the photon is emitted in a superposition of all three directions and the amplitude in each of them is different [see trace E in Fig.~\ref{fig: directions}(b)]. A more thorough analysis can be found in Appendix~\ref{app: direction_4atom}.

\section{\label{sec: coherentInteractions} Long-lived coherent interactions between lattice dark states}

Coupling lattice dark states to radiating momentum states results in dissipative dynamics between both states (as a decay path through the radiating state is always available) that can be used to store excitations in the lattice or to control the properties of the emitted light. One can further exploit the suppressed decay rate of dark states by coupling them with one another and thus engineer extremely long-lived coherent interactions between them. Although the atomic excitation corresponding to a dark state is delocalized and shared among all the dipoles of the lattice, we show in this section that their dynamics are similar to those of coupled individual quantum emitters \cite{Taylor_controllinginteractions,Ana_controllinginteractions}. 

The simplest type of interaction is obtained in the case of two coupled dark states, which can be achieved by a spatial detuning pattern with a period of two atoms. Let us denote their amplitudes in the infinite lattice case as $v_1$ and $v_2$ and their corresponding energy shifts as $J_1$ and $J_2$. In the case of constant detuning, the equations of motion $\dot{v}_1=-iJ_1 v_1 + i \Delta v_2$ and $\dot{v}_2=-iJ_2 v_2 + i \Delta v_1$ generate Rabi oscillations between the dark states. If the excitation is initially stored in $v_1$, the amplitude in the other momentum state takes the simple form

\begin{eqnarray}
    v_2(t)= i \frac{\Delta}{\Omega_{\textit{gen}}} e^{-i \frac{J_1+J_2}{2}t} \sin (\Omega_{\textit{gen}} t),
\end{eqnarray}

\noindent where $\Omega_{\textit{gen}}=\sqrt{{\Delta^2 + \left(\frac{J_1-J_2}{2}\right)^2}}$ is the generalized Rabi frequency.

The oscillation frequency of the momentum pair as well as their global phase depend on the difference between energy shifts $ J = J_1-J_2$. For a stored excitation with a momentum distribution of finite width, this difference will vary among momentum pairs and a distribution of $J$ values appears, which can cause two types of dephasing. The first one comes in through the phase $\exp(-i(J_1+J_2)t/2)$ and therefore 
only introduces relative phases between momentum states. Such a dephasing would already happen without the detuning pattern and does not modify the direction of the photon emitted through a later retrieval scheme. 
The second source of decoherence originates from the difference in Rabi frequencies $\Omega_{\textit{gen}}$ and results in different momentum pairs undergoing Rabi oscillations with slightly different frequencies. One can estimate the time $t_{\textit{\textit{dephase}}}$ needed for two momentum pairs to have their maxima displaced by a certain angle $\Phi_{\textit{min}}$ as $\Omega_{gen} (\mathbf{k}_c) t_{\textit{dephase}} =  \Omega_{gen} (\mathbf{k}\neq \mathbf{k}_c)t_{\textit{dephase}} \pm \Phi_{\textit{min}}$, where $\mathbf{k}_c$ is the central momentum of the original distribution and $\mathbf{k}$ is a momentum in the neighborhood of $\mathbf{k}_c$. Dividing this quantity by the period of the oscillations $T \approx \pi/\Omega_{gen} (\mathbf{k}_c)$, one can obtain the quality factor of the oscillations between two momentum pairs:

\begin{eqnarray}
\label{eq: qualityfactor}
   Q^{(\textit{dephase})} (\mathbf{k})\approx \frac{t_{\textit{dephase}}}{T} \approx \frac{\Omega_{gen} (\mathbf{k}_c) \Phi_{\textit{min}}/\pi}{|\Omega_{gen} (\mathbf{k}_c)-\Omega_{gen} (\mathbf{k})|}.
\end{eqnarray}

\begin{figure}[b]
\includegraphics[width=\linewidth]{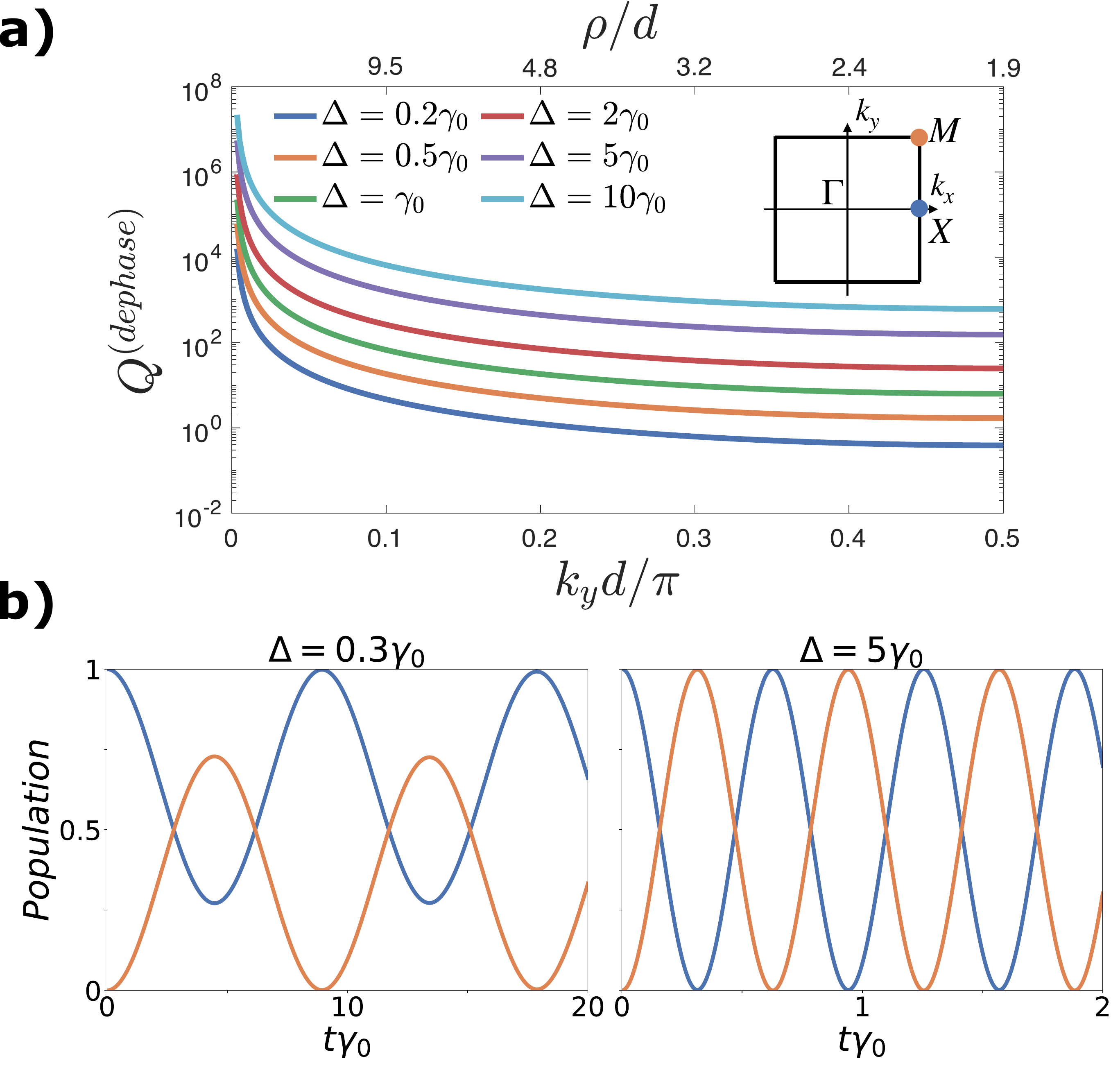}
\caption{\label{fig: rabi} \textbf{Rabi flopping of two momentum dark states.} (a) Quality factor of the coherent oscillations, limited by decoherence effects arising from a finite momentum width of the dark state stored at the $X$ point of the Brillouin zone. The quality factor is plotted versus momentum in reciprocal space (lower $x$ axis) or, analogously, versus the beam waist such that $99.8\%$ of the excitation lies within the corresponding momentum (upper axis). The inset shows how the detuning profile $\Delta_R=\Delta (-1)^{n_y}$ couples momentum states $X$ (blue) and $M$ (orange). (b) Rabi oscillations between momentum states with spread $\rho=1.8\lambda_0$ centered around $X$ (blue trace) and $M$ (orange trace) for $\Delta=0.3\gamma_0$ and $\Delta=5\gamma_0$ in a $21 \times 21$ lattice of spacing $d=0.3\lambda_0$. } 
\end{figure}

In Fig.~\ref{fig: rabi}, we show the decoherence quality factor for an infinite square lattice of spacing $d=0.3\lambda_0$ and such that the initial collective excitation centered at the Brillouin point $X$ is coupled to $M$ via the detuning pattern $\Delta_j=\Delta (-1)^{n_y}$.  The quality factor increases with increasing $\Delta$ and is lower for momenta farther away from the center frequency $\mathbf{k}_c$. On the $x$ axis at the top of the figure, we show the waist of the Gaussian photon such that $99.8\%$ of the excitation is contained within the corresponding momentum range. For example, for a waist $\rho=6d=1.8\lambda_0$ and a detuning $\Delta=10\gamma_0$, $99.8\%$ of the excitation has a quality factor $\sim 5 \times 10^3$. Because the predominant momenta will be around $\mathbf{k}_c$, this is only a very conservative estimate and we can generally expect quality factors for Gaussian momentum distributions to be larger than those given by Eq.~(\ref{eq: qualityfactor}). Figure~\ref{fig: rabi}(b) shows two examples of the resulting long-lived Rabi oscillations between momentum states centered at $X$ (blue) and $M$ (orange) for a Gaussian photon of waist $\rho=6d$ stored in a $21 \times 21$ lattice of spacing $d=0.3\lambda_0$. Note that the oscillations can be experimentally measured by changing the detuning pattern and coupling the $X$ point (or $M$ point) to the $\Gamma$ point, forcing the excitation in that momentum state to leave the atomic array in the perpendicular direction. 

\begin{figure}[b]
\includegraphics[scale=0.35]{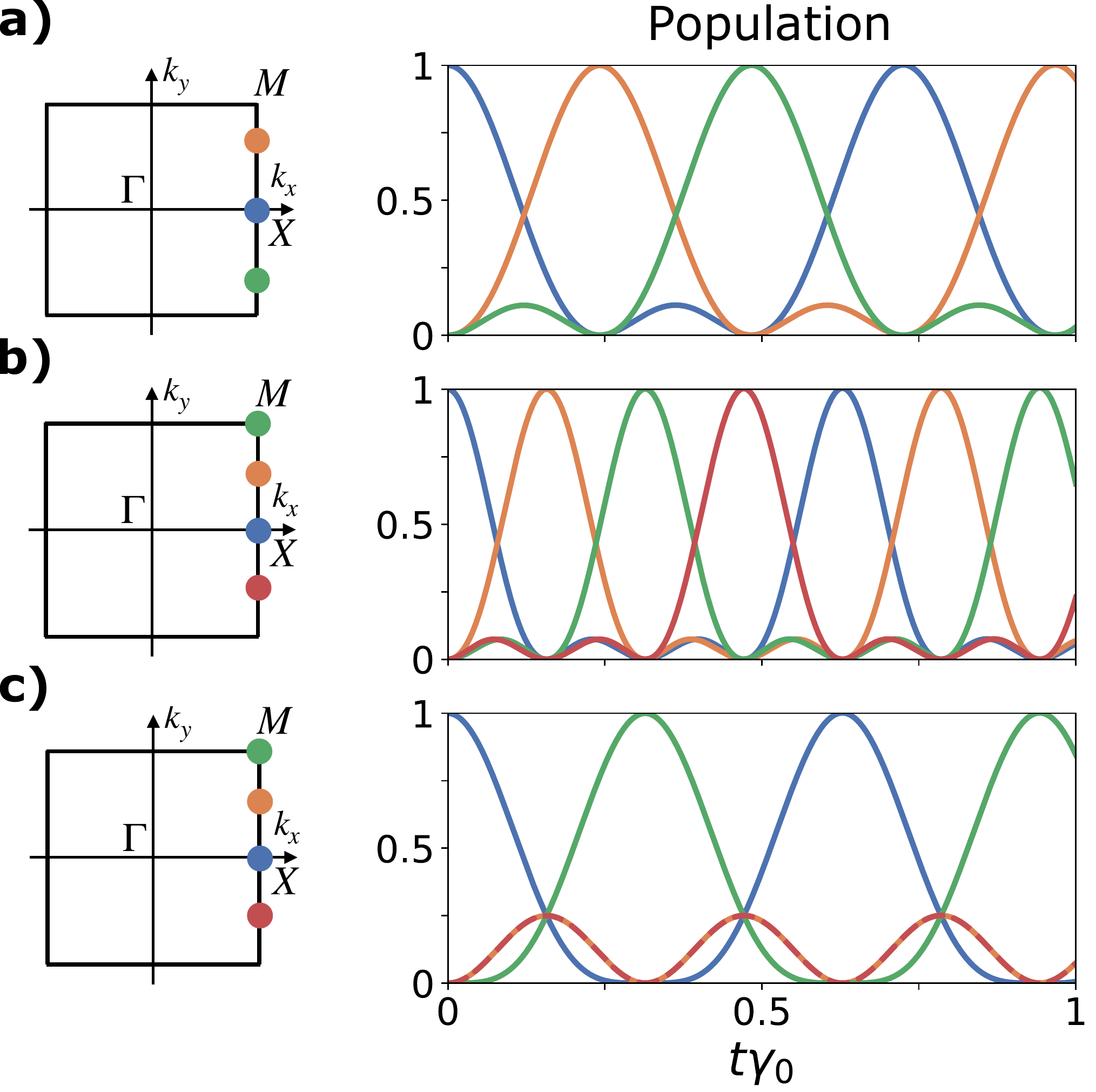}
\caption{\label{fig: multi_osc} \textbf{Engineering coherent interactions.} Cyclic population transfer between (a) three and (b) four lattice dark states and (c) alternative four-atom dynamics with partially suppressed population at states $\mathbf{k}=(\pi,\pm \pi/2)$ (orange and red). The different detuning patterns have the following Fourier components: (a) $\mathbf{Q}_{\pm}=(\pm 2\pi/3d,0)$ and (c) $\tilde{\mathbf{Q}}_{\pm}=(\pm \pi/2d,0)$ with amplitudes $\pm 5i$; (b) $\tilde{\mathbf{Q}}_{\pm}$ with amplitudes $5\sqrt{2}e^{\pm i \pi/4}$ and $\mathbf{Q}_{\pi}=( \pi/d,0)$ with amplitude $5$.
High-quality coherent interactions between multiple dark states require large arrays and narrow excitation distributions in reciprocal space. Here, we consider a $61 \times 61$ lattice with spacing $d=0.4\lambda_0$ in (a) and $d=0.3\lambda_0$ in (b) and (c), and incident photons with waists $\rho=16d$.
}
\end{figure}

It is possible to engineer more complex interactions between a larger number of momentum states by applying detuning patterns with larger spatial periods. In general, a detuning landscape with $m$ atoms per unit cell couples $m$ different momentum states. By properly choosing the weights of each Fourier component $\mathbf{Q}$ of the detuning, one can control the specific nature of the coupling and achieve a wide variety of dynamics. For example, the symmetry of the detuning patterns  $\Delta_A$ and $\Delta_D$ presented in Sec.~\ref{sec: direction} reduces the dynamics of the system to those of simple Rabi oscillations between superpositions of dark momentum states. More interestingly, one can produce interactions that create a cyclic population transfer between momentum states $\mathbf{k}=(\pi,0) \rightarrow (\pi,\pi/n) \rightarrow (\pi,2\pi/n) \rightarrow ... \rightarrow (\pi,(n-1)\pi/n)=(\pi,-\pi/n)$. In Fig.~\ref{fig: multi_osc}(a) and ~\ref{fig: multi_osc}(b), we show such cyclic dynamics between three and four momentum states, respectively, as well as the detuning patterns used to achieve them. Alternatively, it is possible to partially suppress the amplitude in some momentum states through interference effects. This is the case of Fig.~\ref{fig: multi_osc}(c), where the momentum states $\mathbf{k}=(\pi,0)$ and $\mathbf{k}=(\pi,\pi)$ oscillate between zero and one, whereas the states $\mathbf{k}=(\pi,\pm \pi/2)$ have identical amplitudes always below $0.25$. 

Again, the excitations can be retrieved by applying an appropriate retrieval pattern. For example, the two-atom pattern $\Delta_{R}=\Delta (-1)^{n_x}$ couples the dark momentum states $(\pi,k_y)$ to $(0,k_y)$. For three-atom patterns, the three momentum states are emitted into different directions of the electromagnetic field provided that $d>\lambda_0/3$. For four-atom patterns, only three momentum states can be retrieved at a time and the threshold is reduced to $d>\lambda_0/4$. This does not only allow to experimentally measure the oscillations and thus the interactions of the system, but also to steer the emitted photon in a similar way as discussed in Sec.~\ref{sec: direction}. Thus, protocols combining oscillations between dark states and more complex retrieval schemes can provide a greater control on the spatial degrees of freedom of the emitted electromagnetic field.

Two-dimensional lattices also emerge as a promising platform for quantum information science. In particular, lattices have been recently proposed to enhance the interaction between individual impurities \cite{Taylor_controllinginteractions,Ana_controllinginteractions} and therefore realize quantum gates \cite{all_optical_1,all_optical_2}. Due to the versatile and nonlinear nature of their interactions, momentum dark states can be envisioned as an alternative platform to realize quantum gates that does not require additional impurity atoms. To engineer high-quality interactions and dynamics, one generally needs well-defined and nonoverlapping excitation distributions. This can be achieved by reducing the lattice spacing, increasing the beam waist of the incident photon, and increasing the number of atoms in the array.

Additionally, coherent interactions between dark momentum states can be used to experimentally measure the bandstructure of atomic arrays in the single-excitation manifold. The frequency of the resulting oscillations and the amplitude of each momentum state solely depend on the energy shifts of each state. Thus, turning on the interactions for different time durations and measuring the state of the emitted photon directly gives the difference in collective energy shifts between different points in reciprocal space. Combining several spatial detuning patterns and incident photons with different angles with respect to the array, one can finally reconstruct its bandstructure.

\section{\label{sec: conclusion} Conclusion and outlook}

In this work, we demonstrate full control over the dark modes that emerge in structured arrays
by applying periodic spatial modulations of the atomic detuning. In quantum optical platforms such as atomic arrays, these detuning profiles can be experimentally achieved by superimposing several optical lattices with varying periodicities, which results in an optical superlattice \cite{Superlattices_1,Superlattices_2}. 

In particular, we show that single photons can be stored in a superposition of dark states and subsequently released with high fidelities for arrays with a few hundred atoms, much less than the size required by other quantum optical platforms such as disordered atomic clouds \cite{Gorshkov_PRL_2007}.
For protocols based on electromagnetically induced transparency that store photons in single-particle dark states \cite{Manzoni_2018,Gorshkov_PRL_2007}, the storage time is given by the lifetime of the third, metastable level. In our case, the light is stored in collective dark states of the lattice that present largely enhanced lifetimes \cite{decay_rate_universal_scaling} and the performance of the memory is limited by the dispersion of the energy bands. The storage time can therefore be significantly improved by considering lattices with alternative geometries that present flat bands. Interestingly, the bandstructure gives a direct handle to modify the momentum profile of the photon. For example, placing the excitation in a region with linear dispersion results in a shift in real space of the emitted electric field. Note that the storage fidelities reported in this work are sensitive to experimental imperfections, such as missing atoms or disorder in the atomic positions. More concretely, the reduction in efficiency is proportional to the fraction of the photon intensity that impinges on the defects \cite{Manzoni_2018} (see Appendix~\ref{app: effect_of_defects}) and to the variance associated with the distribution of atomic displacements \cite{Manzoni_2018, Ephi_2017}.

Additionally, applying detuning patterns with different strengths over time allows to engineer single photons with arbitrary temporal shape and different spectral properties. In particular, one can systematically produce photons in a superposition of multiple frequencies, which can be used as quantum bits in multimode quantum information protocols \cite{controlsinglephoton_2,multidimensionalphotoncontrol_3}. Similarly, combining detuning patterns with different spatial periods allows control of the emission angles of the emitted light field and opens the door to beam steering \cite{beamsteering} at the single-photon level.

These ideas can be extended to atomic arrays beyond the single-excitation manifold \cite{Ana_controllinginteractions,morenocardoner_twophotons,valentin_rydberg,Jen_multiphoton,Molmer_multiexcitation}. In particular, the potential of ordered arrays to store multiple photons is still unknown and could help envision atomic arrays as quantum metasurfaces capable of generating entanglement between different photons and producing more complex quantum states of light \cite{Bekenstein2020,metasurfaces}. Additionally, the long-lived nature of lattice dark states could ease the requirements on the interaction strengths needed to create phase shifts between different photons, relevant to produce quantum gates \cite{morenocardoner_twophotons} and study correlations and many-body physics. In that regard, accessing dark states in the multiple-excitation regime and engineering interactions between them may allow to devise more diverse quantum gates and protocols than the ones possible with a single excitation.

From a fundamental point of view, the combination of atomic arrays and spatial modulations of the atomic detuning can open the door to studying the nature and properties of collective subradiant states. As opposed to current experimental methods with atomic clouds, which rely on waiting for a small fraction of the system to decay into subradiant states \cite{Browaeys_subradiance,Ana_subradiant_comment}, our protocol provides a direct and easy way to excite them.
Due to their highly directional emission profiles and their properties to store and control light, two-dimensional arrays could also be used as building blocks for quantum-computing architectures in the optical regime \cite{all_optical_1,all_optical_2}. Exploring other types of two-dimensional lattices and arrays of different dimensions could help achieve higher control over the properties of the emitted photons and unveil other quantum functionalities. Recent work by Ballantine and Ruostekoski \cite{roustekoski_2021} has shown how to engineer a Huygens surface with a rectangular bilayer lattice, as well as how to generate entanglement between the atomic array and a cavity. Finally, this work could be extended to other structured systems such as nitrogen-vacancy centers \cite{NVcentres} and excitons in atomically thin semiconductors (i.e., transition metal dichalcogenides \cite{exciton_1,exciton_2,trond_andersen}), where superlattices can be optically imprinted to modify the properties of the bandstructure \cite{exciton_superlattice,trond_andersen}.

\begin{acknowledgments}
This work has been supported by the NSF through the CUA Physics Frontier Fund (for partial funding of O.R.B.), and through Grant No. PHY-1912607 (partially funding S.F.Y., regarding basic formalism), the AFOSR through Grant No. FA9550-19-1-0233 (partially funding S.F.Y., quantum information applications), and the DOE through DE-SC0020115 (partially funding S.F.Y., photon manipulating aspects).
O.R.B. acknowledges support from Fundació Bancaria “la Caixa” (LCF/BQ/AA18/11680093). V.W. acknowledges support from the NSF through a grant for the Institute for Theoretical Atomic, Molecular, and Optical Physics at Harvard University and the Smithsonian Astrophysical Observatory.
\end{acknowledgments}

\nocite{*}

\appendix

\section{Green's function}

The dyadic Green's function in free space used in Eq.~(\ref{eq: shifts_greensfunction}) can be written in Cartesian coordinates as \cite{Chew_dyadicGreens,dyadic_novotny_hecht_2006}

\begin{align}
G_{\alpha \beta} &= -\frac{e^{i k r}}{4\pi r} \left[ \left( 1 + \frac{i}{kr} - \frac{1}{(kr)^2} \right) \delta_{\alpha\beta} \right. \nonumber \\
  &+ \left. \left(-1 - \frac{3i}{kr} + \frac{3}{(kr)^2} \right) \frac{r_\alpha r_\beta}{r^2} \right] + \frac{\delta_{\alpha \beta} \delta^{(3)}(\mathbf{r})}{3k^2},
\end{align}

\noindent where $k=\omega/c$, $r=\sqrt{x^2+y^2+z^2}$, and $\alpha,\beta=x,y,z$.

\section{Bandstructure of a non-Bravais lattice}

For an infinite lattice with two atoms per unit cell, as it is the case of the checkerboard lattice presented in Sec.~\ref{subsec: checkerboard}, the two bands can be obtained by diagonalizing the $2 \times 2$ matrix $\mathbf{M}$ for each quasimomentum $\mathbf{k}$ in the first Brillouin zone \cite{Janos_2017,Janos_2017_PRL}. The components of $\mathbf{M}$ are

\begin{equation}
     M_{\mu \nu}=(\omega_0^{(1)} - i \Gamma_0/2) \delta_{1\mu} \delta_{1\nu}+ (\omega_0^{(2)} - i \Gamma_0/2) \delta_{2\mu} \delta_{2\nu} + \chi_{\mu \nu},
\end{equation}

\noindent where $\mu,\nu=1,2$ represent each of the two sublattices and $\omega_0^{(1)}$ and $\omega_0^{(2)}$ are their corresponding transition frequencies. The last term describes atom-atom interactions and is given by

\begin{align}
    \chi_{\mu \nu} = \frac{3 \pi \Gamma_0 c}{\omega_0^{(1)}} & \left[ \sum_{\mathbf{R}_1\neq 0} e^{i \mathbf{k} \mathbf{R}_1} G(\mathbf{R}_1) \delta_{1\mu} \delta_{1\nu}  \right. \nonumber \\
    & + \sum_{\mathbf{R}_1} e^{i \mathbf{k} \mathbf{R}_1} G(\mathbf{R}_1+\mathbf{b}) \delta_{1\mu} \delta_{2\nu} \nonumber \\
    &+ \sum_{\mathbf{R}_2\neq 0} e^{i \mathbf{k} \mathbf{R}_2} G(\mathbf{R}_2) \delta_{2\mu} \delta_{2\nu} \nonumber \\
    &+ \left. \sum_{\mathbf{R}_2} e^{i \mathbf{k} \mathbf{R}_2} G(\mathbf{R}_2-\mathbf{b}) \delta_{2\mu} \delta_{1\nu} \right] , 
\end{align}

\noindent where $\{ \mathbf{R}_1\}$ and $\{ \mathbf{R}_2\}$ are the atomic positions in each sublattice. Additionally, we have assumed that the difference in transition frequencies is small such that $1/\omega_0^{(1)} \approx 1/\omega_0^{(2)}$.

For a non-Bravais lattice with more than two atoms per unit cells, similar expressions can be obtained. In that case, the dimension of $\mathbf{M}$ is $m \times m$.

\section{Populating lattice dark states with a low-intensity, classical driving field}
\label{app: classicalDrive}

In Sec.~\ref{sec:introduction}, we presented the Hamiltonian of the system and the equations of motion in the absence of an external field. If we now consider a weak external drive such that only the one-excitation sector is relevant \cite{Ephi_2017}, the effective Hamiltonian in Eq.~(\ref{eq: efectiveHamiltonian}) acquires the extra term $\sum_j \Omega_j \left( \sigma_j^\dagger + \sigma_j \right)$, where $\Omega_j$ is the spatially dependent Rabi frequency. The resulting equations of motion are

\begin{align}
        \frac{d e_j}{dt} & = i \Delta_j(t) e_j - i \sum_{i} \left( J_{ji} -i \frac{\Gamma_{ji}}{2} \right) e_i
        + i \Omega_j g, \nonumber \\
        \frac{d g}{dt} & = i \sum_i \Omega_i e_i.
\end{align}

For an infinite lattice, the system is described by the amplitudes in each quasimomentum state within the first Brillouin zone. In the case of a checkerboard detuning pattern, we have

\begin{equation}
    \frac{d v_{\mathbf{k}}}{dt}=  - i \left( J_{\mathbf{k}} - i \frac{\Gamma_\mathbf{k}}{2} \right) v_{\mathbf{k}} + i \Delta  v_{\mathbf{k}+\boldsymbol{\pi}/d} + i \Omega_\mathbf{k} g.
\end{equation}

In the weak-driving limit, we can approximate $g \approx 1$. A radiating state $\mathbf{k}_r$ inside the light cone is coupled to the dark state $\mathbf{k}_d=\mathbf{k}_r+\boldsymbol{\pi}/d$ and the equations of motion for the weakly driven system are

\begin{align}
        \frac{d v_{\mathbf{k}_r}}{dt } & =  - i \left( J_{\mathbf{k}_r} - i \frac{\Gamma_{\mathbf{k}_r}}{2} \right) v_{\mathbf{k}_r} + i \Delta  v_{\mathbf{k}_d} + i \Omega_{\mathbf{k}_r}, \nonumber \\
        \frac{d v_{\mathbf{k}_d}}{dt} & =  - i J_{\mathbf{k}_d} v_{\mathbf{k}_d} + i \Delta  v_{\mathbf{k}_r},
\end{align}

\noindent and an incoming drive with a certain in-plane momentum excites lattice components with the same quasimomentum. As a result, the dark state is not driven by the external field, as its quasimomentum lies outside the light cone. Instead, it is populated through the coupling with the radiating state produced by the spatial detuning pattern.

Starting with all the atoms in the ground state, we obtain the steady-state amplitude in the dark state,

\begin{equation}
\label{eq: InitialdarkstateProfile}
    v_{\mathbf{k}_d}= -\frac{\Delta \Omega_{{\mathbf{k}_r}}}{\Delta^2 - \left( J_{\mathbf{k}_r} - i \frac{\Gamma_{\mathbf{k}_r}}{2} \right)J_{\mathbf{k}_d} }.
\end{equation}

For an incident Gaussian beam perpendicular to the array, with waist $\rho$ and whose focal plane coincides with the position of the lattice $z=0$, the electric field distribution in the lattice plane is $\mathbf{E}(x,y)=E_0 \boldsymbol{\epsilon}_G e^{-(x^2+y^2)/\rho^2}$, which corresponds to a Gaussian distribution $\mathcal{N}(0,\rho^2)$. The weights $\Omega_{{\mathbf{k}_r}}$ are proportional to the momentum-space distribution of the electric field, which is again a Gaussian distribution $\mathcal{N}(0,\rho^{-2})$. From the band structure in Fig.~\ref{fig: schematic}(c), one can see that the term $\left( J_{\mathbf{k}_r} - i \Gamma_{\mathbf{k}_r} /2 \right)J_{\mathbf{k}_d}$ varies only weakly around the $\Gamma$ and $M$ points. A Gaussian classical drive therefore populates a near-Gaussian distribution of dark momentum states. 

In Sec.~\ref{sec: memory}, we study the retrieval process of dark states given by Eq.~(\ref{eq: InitialdarkstateProfile}) for different values of beam waist $\rho$ and show that they can be retrieved into Gaussian modes of the same waist with almost unit efficiency.

\section{Frequency modulation}
\label{app: frequency modulation}

For a lattice with a constant checkerboard detuning pattern of magnitude $\Delta$ and a time-dependent homogenous detuning $\xi(t)$ equal at all lattice sites, the equations of motion for the amplitudes in the dark $v_d$ and radiating $v_r$ momentum states are given by

\begin{eqnarray}
    \dot{v}_d=-i J_{d} v_d + i \Delta(t) v_r \nonumber + i \xi(t) v_d, \\
    \dot{v}_r=-i \left( J_{r} -\Gamma_r/2 \right) v_r + i \Delta(t) v_d + i \xi(t) v_r.
\end{eqnarray}

If the dynamics for $\xi(t)=0$ are given by $v_d^{(0)}(t)$ and $v_r^{(0)}(t)$, the solution for a general $\xi$ is simply given by

\begin{eqnarray}
    v_d(t)=v_d^{(0)}(t) e^{i \zeta(t)}, \nonumber\\
    v_r(t)=v_r^{(0)}(t) e^{i \zeta(t)}, \nonumber\\
\end{eqnarray}

 with $\zeta(t) = \int_0^t \xi(\tau) d\tau$.
 
 If $\xi(t)$ is additionally periodic, such that $\xi (t+2\pi/\Omega) = \xi(t) $, the term $e^{i \zeta(t)}$ can be expanded as \cite{Silveri_2017_modulation}
 
 \begin{eqnarray}
    A(t)=e^{i\zeta(t)}= \sum_{n=-\infty}^{\infty} a_n e^{in\Omega t}, \nonumber \\
    a_n=\frac{\Omega}{2\pi} \int_{0}^{2 \pi/\Omega} e^{-in\Omega \tau} e^{i \zeta(\tau)} d\tau .
\end{eqnarray}

For a sinusoidal modulation $\xi(t)= \delta \cos (\Omega t)$, these coefficients are $a_n=J_n(\delta/\Omega)$, where $J_n$ is the Bessel function of order $n$ \cite{Silveri_2017_modulation}. The amplitude in the radiating field is then

\begin{align}
    v_r(t) & =\sum_{n=-\infty}^\infty J_n ( \delta/\Omega ) v_r^{(0)}(t) e^{in \Omega t} \nonumber  \\
    & =\sum_{n=-\infty}^\infty  \frac{\Delta J_n ( \delta/\Omega )}{2\sqrt{\Delta^2+G^2}} \left(e^{i (\omega_+ + n \Omega) t} -e^{i (\omega_- +n \Omega) t} \right) .
\end{align}

The frequency profile of the outgoing photon with momentum matching that of the radiating state can be obtained from Eq.~(\ref{eq: field_infinite}):

\begin{align}
    \mathbf{E}_{\boldsymbol{\kappa}_{||}=\mathbf{k}_r}(\omega) & \propto   \sum_{n=-\infty}^\infty  J_n ( \delta/\Omega) \times \nonumber \\
    & \left( \frac{1}{\omega - \omega_0 + \omega_+ + n \Omega}   -  \frac{1}{\omega - \omega_0 + \omega_- + n \Omega} \right) 
\end{align}

\begin{figure}[t]
\includegraphics[width=\linewidth]{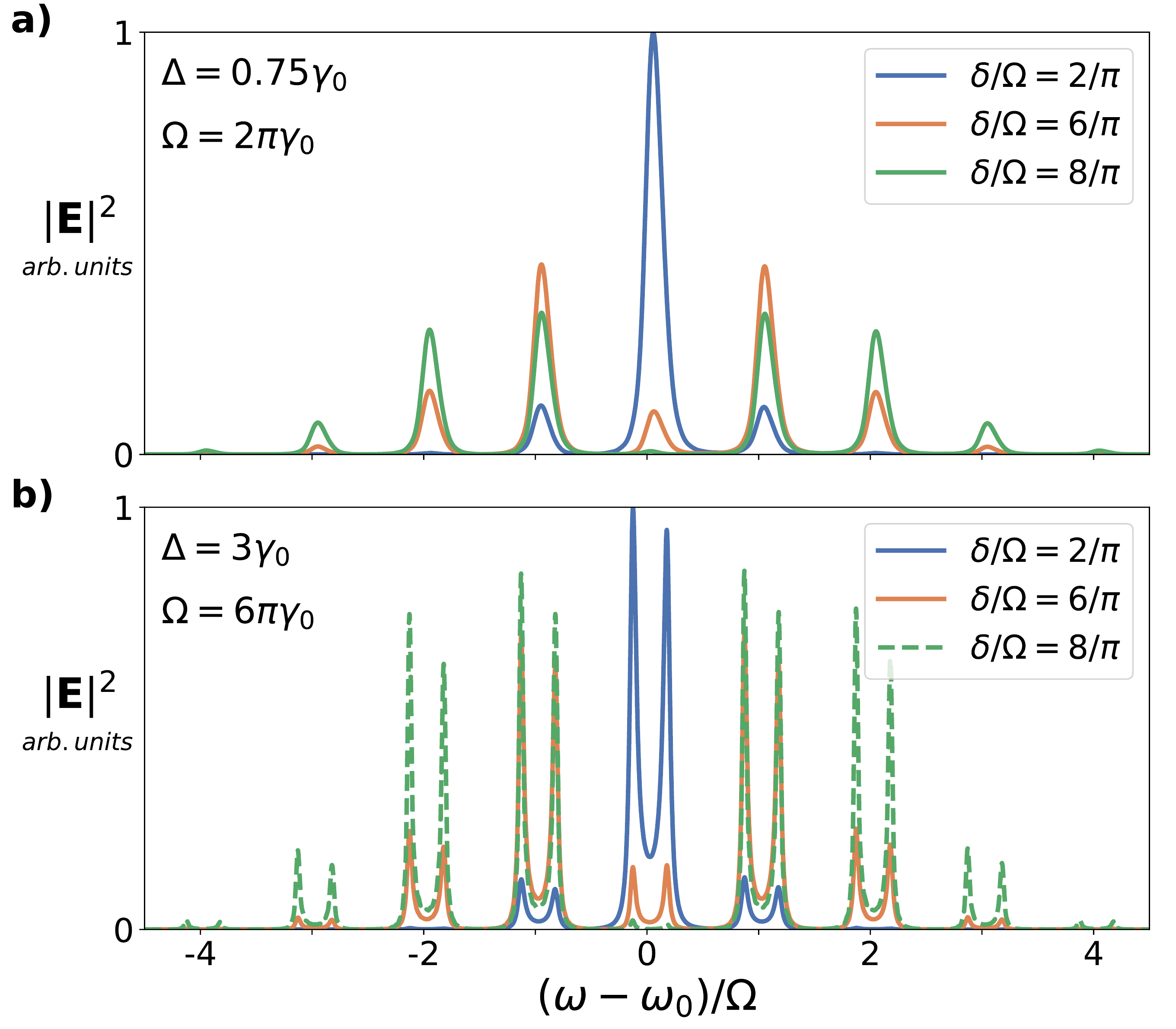}
\caption{\label{fig: freq_modulation} \textbf{Frequency modulation of emitted photon.} Frequency profile of the emitted electromagnetic field from a $21 \times 21$ lattice with spacing $d=0.3\lambda_0$ and an initial Gaussian state of waist $\rho=6d$. (a) A checkerboard detuning pattern with strength $\Delta=0.75\gamma_0$ is applied and the frequency of the homogeneous detuning is $\Omega=2\pi \gamma_0$. (b) The parameters are $\Delta=3\gamma_0$ and $\Omega=6\pi \gamma_0$. The three traces correspond to different ratios of the relevant parameter $\delta/\Omega$.}
\end{figure}

Comparing with the result in Eq.~(\ref{eq: freq_profile}), the modulation $\xi(t)$ just replicates the original profile every $n\Omega$. For a good resolution of the sidebands, one requires $\Omega$ to be larger than the width of the original profile, $\sim 2\Delta+\Gamma_r$. The amplitude of each sideband $n$ can be controlled by adjusting the ratio $\delta/\Omega$. Figure~\ref{fig: freq_modulation} shows the resulting frequency profiles for different combinations of $\delta/\Omega$ and for two different detuning amplitudes $\Delta=1$ and $\Delta=4$. In the first case, a single peak is replicated, whereas in the second case the repeating unit contains two peaks.

\section{Direction of emitted photon for detuning pattern with a four-atom period}
\label{app: direction_4atom}

Let us consider an excitation initially stored at $\mathbf{k}=(\pi/d,0)$. The most general detuning pattern with four atoms in the unit cell and along the $x$ axis contains the Fourier components $\mathbf{Q}_0=0$, $\mathbf{Q}_\pm=(\pm \pi/2d,0)$, and $\mathbf{Q}_\pi=(\pm \pi/d,0)$ and can be written as $\Delta_j= \alpha e^{i \mathbf{Q}_0 \mathbf{r}_j} + \beta e^{i \mathbf{Q}_+ \mathbf{r}_j} + \beta^* e^{i \mathbf{Q}_- \mathbf{r}_j} + \delta e^{i \mathbf{Q}_\pi- \mathbf{r}_j}$ such that $\alpha,\delta \in \mathbb{R}$ and $\beta \in \mathbb{C}$. It couples the momentum states $\mathbf{k}=(\pi/d,0)$ with $\mathbf{k}=(\pm \pi/2d,0)$ and $\mathbf{k}=\mathbf{0}$. The photon will in general be emitted in a superposition of three different directions: the axis perpendicular to the array and the two directions forming an angle $\theta=\pm \arcsin( \lambda_0/4d)$ with it. Labeling the states by their generalized momentum $k_x d$ along the $x$ axis, we obtain the equations of motion

\begin{align}
\label{eq: direction_4atomperiod}
        \frac{dv_\pi}{dt}&=-i \left( J_\pi - \alpha \right) v_\pi + i \beta v_{\pi/2} + i \beta^* v_{-\pi/2} + i \delta v_0,  \nonumber \\
        \frac{dv_{\pi/2}}{dt}&=-i \left( J_{\pi/2} -\alpha -i \frac{\Gamma_{\pi/2}}{2} \right) v_{\pi/2} \nonumber \\ &+ i \beta v_{0} + i \beta^* v_{\pi} + i \delta v_{-\pi/2}, \nonumber \\
        \frac{dv_{-\pi/2}}{dt}&=-i \left( J_{\pi/2} -\alpha -i \frac{\Gamma_{\pi/2}}{2} \right) v_{-\pi/2} \nonumber \\ &+ i \beta v_{\pi} + i \beta^* v_{0} + i \delta v_{\pi/2}, \nonumber \\
        \frac{dv_{0}}{dt}&=-i \left( J_{0} -\alpha -i \frac{\Gamma_{0}}{2} \right) v_{0} \nonumber \\ &+ i \beta v_{-\pi/2} + i \beta^* v_{\pi/2} + i \delta v_{\pi}, \nonumber \\
\end{align}

By appropriately choosing the values of each Fourier component, one can produce different types of output photons:

\paragraph{} Choosing for example $\beta \in \mathbb{R}$ and $\alpha=\delta=0$, we obtain equal emission along the directions $\theta=\pm \arcsin( \lambda_0/4d)$, as well as nonzero emission perpendicular to the array. This can be seen by writing Eq.~(\ref{eq: direction_4atomperiod}) in the basis $v_{\pm}=v_{\pi/2} \pm v_{-\pi/2}$

\begin{align}
        \frac{dv_\pi}{dt}&=-i J_\pi v_\pi + i \beta v_+, \nonumber \\
        \frac{dv_+}{dt}&=-i \left( J_{\pi/2} -i \frac{\Gamma_{\pi/2}}{2} \right) v_{+} + 2i \beta (v_\pi + v_0),   \nonumber \\
        \frac{dv_{0}}{dt}&=-i \left( J_{0} -i \frac{\Gamma_{0}}{2} \right) v_{0} + i \beta v_+, \nonumber \\
        \frac{dv_-}{dt}&=-i \left( J_{\pi/2} -i \frac{\Gamma_{\pi/2}}{2} \right) v_{-} .
\end{align}

\noindent For an initial excitation in the dark state $v_\pi$, only the symmetric superposition $v_+$ will be populated. Thus, $v_-(t)=0$, such that $v_{\pi/2}=v_{-\pi/2}$ during the whole time evolution and emission along both nonorthogonal directions is identical. Additionally, $v_0$ is coupled to $v_+$ and will be populated during the decay process, such that emission perpendicular to the array will also occur. The relative amplitudes between perpendicular and oblique emission can be controlled through the parameter $\beta$.

\paragraph{} The emission perpendicular to the array can be suppressed by choosing a proper complex $\beta=|\beta|e^{i\pi/4}$. In that case, the momentum states $v_{\pm \pi/2}$ destructively interfere and the amplitude at $v_0$ vanishes. In the basis $\tilde{v}_{\pm}=e^{i\pi/4}v_{\pi/2} \pm e^{-i\pi/4}v_{-\pi/2}$, the equations of motion are 

\begin{align}
        \frac{dv_\pi}{dt}&=-i J_\pi v_\pi + i |\beta| \tilde{v}_+, \nonumber \\
        \frac{d\tilde{v}_+}{dt}&=-i \left( J_{\pi/2} -i \frac{\Gamma_{\pi/2}}{2} \right) v_{+} + 2i |\beta| v_\pi,   \nonumber \\
        \frac{dv_{0}}{dt}&=-i \left( J_{0} -i \frac{\Gamma_{0}}{2} \right) v_{0} + |\beta| \tilde{v}_-, \nonumber \\
        \frac{d\tilde{v}_-}{dt}&=-i \left( J_{\pi/2} -i \frac{\Gamma_{\pi/2}}{2} \right) \tilde{v}_{-} - 2 |\beta| v_0 .
\end{align}

\noindent Now, the dynamics are split in two uncoupled blocks: one that contains the dark state $v_\pi$ and the radiating superposition $\tilde{v}_+$, and a second one that comprises the radiating states $\tilde{v}_-$ and $v_0$. For an excitation originally stored in the dark state, we have $\tilde{v}_-(t)=v_0(t)=0$. The resulting photon therefore has a suppressed amplitude along the perpendicular direction $\theta=0$ and equal amplitudes in the other two oblique directions.

\paragraph{} Keeping the complex $\beta=|\beta|e^{i\pi/4}$ and introducing a finite $\delta$, one couples the states $v_\pi$ and $v_0$. As a result, the two blocks get coupled and it is no longer true that $|v_{\pi/2}|=|v_{-\pi/2}|$. The photon is emitted in a coherent superposition of all three directions, such that the amplitudes in all directions are in general different and determined by the relative strengths of $|\beta|$ and $\delta$.

An example of the magnitude of the electric field as a function of propagation direction for all three cases can be found in Fig.~\ref{fig: directions}(b) of the main text.

\section{Effect of defects in the lattice}
\label{app: effect_of_defects}

\begin{figure}
\includegraphics[scale=0.32]{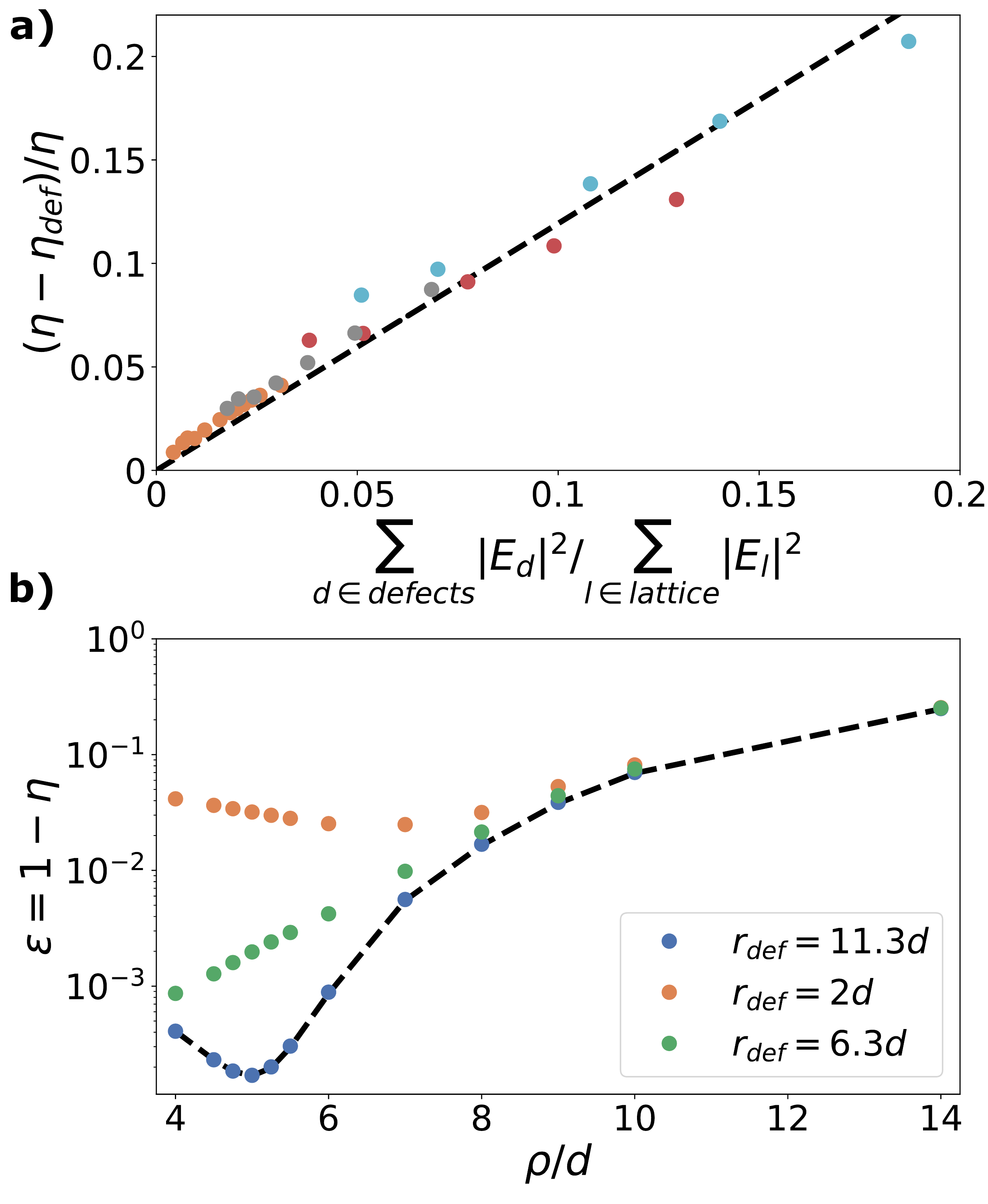}
\caption{\label{fig: effect_defect} \textbf{Effect of defects.} (a) Relative decrease in efficiency $(\eta -\eta_{def})/\eta$ as a function of $\sum_{d \in \textit{defects}} |E_d|^2 / \sum_{l \in \textit{lattice}} |E_l|^2$. A linear relation with proportionality constant $\alpha=1.19$ is found between both quantities (black dashed line). The different colors represent results for systems with different sets of holes: orange has a hole at $(2d,0)$; grey has three holes at $(2d,0)$, $(-d,-d)$ and $(6d,3d)$; red corresponds to a set of seven holes at $(0,0)$, $(2d,0)$, $(-2d,2d)$, $(-3d,d)$, $(5d,-d)$, $(4d,4d)$ and $(8d,8d)$; cyan corresponds to a set of nine holes, the seven in red plus $(-d,d)$ and $(0,-3d)$.  The origin of the lattice is considered to be at $(0,0)$. b) Infidelity versus beam waist $\rho$ for different single-atom defects. The blue dots corresponds to a defect at position $(8d,8d)$, the green dots to a defect at $(4d,3d)$ and the orange to a defect at $(2d,0)$. The black dashed line is the result for the corresponding finite lattice without defects. In all cases, a lattice with $d=0.3\lambda$ and $21 \times 21$ atoms is considered.}
\end{figure}

The performance of the array as a memory, as well as its potential to modify the properties of single photons, are generally affected by imperfections of the system. Here, we study the role of missing atoms in the array, which we refer to as defects. Let us define the storage efficiency for a finite array of atoms as $\eta$ and the efficiency for a lattice with a set of defects as $\eta_{def}$. Figure~\ref{fig: effect_defect}a shows that the relative decrease in efficiency $(\eta-\eta_{def})/\eta$ is proportional to the fraction of the incoming intensity that impinges on the missing atoms. One can therefore write \cite{Manzoni_2018}

\begin{equation}
    \eta_{def} \approx  \eta \left( 1 - \alpha \frac{\sum_{d \in \textit{defects}} |E_d|^2}{\sum_{l \in \textit{lattice}} |E_l|^2} \right),
\end{equation}

\noindent where $\alpha$ is a constant, $E_j$ is the amplitude of the detection mode at the atomic position $\mathbf{r}_j$, and the sums in the denominator and the numerator run over the positions of the whole (defectless) lattice and the positions of the defects, respectively.

As a result, the efficiency is mostly unaffected by defects lying far from the center of the array (assuming that the incoming photon is well centered), as shown by the blue curve in Fig.~\ref{fig: effect_defect}(b). However, the efficiency can be considerably reduced by defects lying close to the center of the array. As demonstrated by the orange trace, the efficiency of a $21 \times 21$ lattice with spacing $d=0.3\lambda_0$ and a hole at a distance $2d$ from the center of an incoming photon with waist $\rho=4d$ is reduced to $\sim 96\%$. The approximate $4\%$ decrease is close to the percentage of the incoming light that hits the defect, which amounts to $\sim 3\%$. Note also that the decay in efficiency gets suppressed for large enough beam waists, when the ratio of intensities between a single defect and the perfect lattice with $N$ atoms approaches $1/N$.

\bibliography{apssamp}

\end{document}